\documentclass[12pt,preprint]{aastex}

\usepackage{emulateapj5}

\newcommand{\kms}{\,km\,s$^{-1}$}

\newcommand{\arcs}{$^{\prime\prime}$}

\newcommand{\fuse}{{\em FUSE}}
\newcommand{\iue}{{\em IUE}}
\newcommand{\imaps}{{\em IMAPS}}
\newcommand{\hst}{{\em HST}}
\newcommand{\copernicus}{{\em Copernicus}}

\shortauthors{Hoopes et al.}
\shorttitle{Deuterium Toward HD~195965 and HD~191877}

\begin{document}


\title{Deuterium Toward Two Milky Way Disk Stars: Probing Extended Sight Lines with the Far Ultraviolet Spectroscopic Explorer}

\author{Charles G. Hoopes\altaffilmark{1}, Kenneth R. Sembach\altaffilmark{2}, Guillaume H\'ebrard\altaffilmark{3}, H. Warren Moos\altaffilmark{1}, \\ and David C. Knauth\altaffilmark{1}}
\altaffiltext{1}{Department of Physics and Astronomy, Johns Hopkins University, 3400 North Charles Street, Baltimore, MD 21218; choopes@pha.jhu.edu, hwm@pha.jhu.edu, dknauth@pha.jhu.edu}
\altaffiltext{2}{Space Telescope Science Institute, 3700 San Martin Drive, Baltimore, MD 21218; sembach@stsci.edu}
\altaffiltext{3}{Institut d'Astrophysique de Paris, 98 bis Boulevard Arago, 75014 Paris, France; hebrard@iap.fr}

\begin{abstract}

We have carried out an investigation of the abundance of deuterium
along two extended sight lines through the interstellar medium (ISM)
of the Galactic disk. The data include {\it Far Ultraviolet
Spectroscopic Explorer (FUSE)} observations of HD~195965 (B1Ib) and
HD~191877 (B0V), as well as Space Telescope Imaging Spectrograph
(STIS) observations of HD~195965. The distances to HD~195965 and
HD~191877, derived from spectroscopic parallax, are $794\pm200$~pc and
$2200\pm550$~pc, respectively, making these the longest Galactic disk
sight lines in which deuterium has been investigated with \fuse. The
\fuse\ spectra contain all of the \ion{H}{1} Lyman series
transitions (and the corresponding D transitions) except
Ly$\alpha$. The higher Lyman lines clearly show the presence of
deuterium. We use a combination of curve of growth analyses and line
profile fitting to determine the \ion{D}{1} abundance toward each
object.  We also present column densities for \ion{O}{1} and
\ion{N}{1} toward both stars, and \ion{H}{1} measured from Ly$\alpha$
absorption in the STIS spectrum of HD~195965. Toward HD~195965 we find
D/H=$(0.85\pm^{0.34}_{0.24})\times10^{-5}$ ($2\sigma$),
O/H=$(6.61\pm^{1.03}_{1.11})\times10^{-4}$, and
N/H=$(7.94\pm^{1.69}_{1.34})\times10^{-5}$. Toward HD~191877 we find
D/H=$(0.78\pm^{0.52}_{0.25})\times10^{-5}$ ($2\sigma$) and
N/H=$(6.76\pm^{2.22}_{1.97})\times10^{-5}$. The \ion{O}{1} column
density toward HD~191877 is very uncertain. Our preferred value gives
O/H=$(3.09\pm^{1.98}_{0.98})\times10^{-4}$, but we cannot rule out O/H
values as low as O/H=$1.86\times10^{-4}$, so the O/H value for this
sight line should be taken with caution. The D/H ratios along these
sight lines are lower than the average value of
($1.52\pm0.15)\times10^{-5}$ ($2\sigma$ in the mean) found with \fuse\
for the local interstellar medium ($\sim 37$ to 179 pc from the
Sun). These observations lend support to earlier detections of
variation in D/H over distances greater than a few hundred pc. The O/H
ratio toward HD~195965 is supersolar. This star is part of an OB
association, so there may be local enrichment by nearby massive
stars. The D/H and O/H values measured along these sight lines support
the expectation that the ISM is not well mixed on distances of
$\sim1000$ pc. These observations demonstrate that although D/H
studies through Lyman absorption may become impractical at d$>$2500 pc
and log $N$(\ion{H}{1})$>$21, D/H studies in the distance range from
500 to 2500 pc may be very useful for investigating mixing and
chemical evolution in the ISM.

\end{abstract}



\section{Introduction}

The abundance of deuterium in the interstellar medium (ISM) is a
strong constraint on chemical evolution scenarios. Unlike most
elements, the net cosmic abundance of deuterium is not enhanced by
stellar processes or supernovae, so its abundance has been declining
from the primordial value due to destruction by stellar
processing. The primordial abundance of deuterium is itself an
important quantity, as it can be used to measure $\eta$, the universal
ratio of baryons to photons, which can be compared to the predictions
of Big Bang nucleosynthesis
\citep{rafs73,bs85,bnt01}.

One approach to finding the initial deuterium abundance is to measure
it in high-redshift QSO absorption-line systems, assuming that the
clouds which give rise to these systems have undergone very little
stellar processing \citep{w97,tblfws99,otkspl01,pb01,ldd02}. However,
the large range in D/H values measured in these systems suggests that
the gas in some of these clouds may in fact have undergone some
stellar processing \citep{pi01,foscv01}. It is therefore crucial to
understand how chemical evolution affects the abundance of deuterium
relative to other elements. Several groups are measuring the abundance
of deuterium and other elements along many sight lines in the Galactic
ISM in order to establish the mean D/H ratio in the Galactic disk, and
to look for variations in the D/H ratio. These variations can shed
light on the effects of star-formation history and mixing processes on
the D/H ratio. The local D/H value can then be compared to regions
that have undergone varying amounts of stellar processing ({\it e.g.,}
high velocity clouds, or intergalactic \ion{H}{1} clouds), which will
illuminate the effects of stellar processing on the D/H ratio in these
systems.

The \copernicus\ satellite was used to measure the \ion{D}{1} and
\ion{H}{1} Lyman series absorption toward $\sim20$ stars, and the results
suggested a D/H ratio of 1.5$\times$10$^{-5}$, with a dispersion that
may reflect real variations in the ISM (see Vidal-Madjar \& Gry 1984
for a review). Several measurements of D/H in the local ISM have been
made with the {\em Hubble Space Telescope} (\hst)
\citep{lin95,l96,lin02,lem96,v98,h99,s99,vpltcg00}, though only the Ly$\alpha$
transitions of \ion{D}{1} and \ion{H}{1} could be analyzed, limiting
the number of available sight lines to those with low enough
\ion{H}{1} columns that the \ion{D}{1} absorption is not blended with
the \ion{H}{1} line but large enough \ion{H}{1} columns that \ion{D}{1}
can be detected. Three sight lines observed with the Interstellar
Medium Absorption Profile Spectrograph (\imaps) spanned larger
distances than the \hst\ sight lines (250 to 500 pc versus $\la$ 100
pc), and like the \copernicus\ results they showed evidence of
variation in D/H with location \citep{j99,stfjvw00}.

One of the science goals of the {\em Far Ultraviolet Spectroscopic
Explorer} (\fuse) is to measure D/H in the Milky Way
\citep{m00}. \fuse\ is uniquely suited to do this because it observes the
spectral region from 905 to 1187~\AA\, which includes all of the
\ion{D}{1} Lyman transitions except Ly$\alpha$. The ability to use
multiple \ion{D}{1} transitions means that \fuse\ is able to measure
$N$(\ion{D}{1}) along higher column density sight lines than missions
that only observe Ly$\alpha$ ({\it e.g.,} \hst), and for many more
sight lines than short term missions ({\it e.g.,} \imaps). Moos et
al. (2002, and references therein) reported results for the first set
of seven sight lines ranging from 37 to 179 pc, five of which had both
\ion{D}{1} measurements (from \fuse) and \ion{H}{1} (from \hst). The
results were consistent with a single value of
D/H$=(1.52\pm0.08)\times10^{-5}$ (the uncertainty is $1\sigma$ in the
mean). However, none of these sight lines extend as far as the three
sight lines observed by \imaps. Analysis of more extended sight lines
is crucial for the verification and understanding of variations in the
D/H ratio in the ISM.

In this paper we present \fuse\ measurements of D/H toward two distant
stars in the Galactic disk: HD~195965 and HD~191877. These stars are
at distances greater than the \imaps\ targets, extending the distance
and range of $N$(\ion{H}{1}) for D/H measurements in the
ISM. Information about these two sight lines is listed in Table 1. The
\fuse, STIS, and ground-based observations and data reduction are
described in \S\ 2. Section 3 describes some properties of the sight
lines. In \S\ 4 we describe both the curve of growth and the
absorption profile fitting analyses of the \ion{D}{1} column density
toward both stars. In \S\ 5 we discuss the determination of the
\ion{O}{1} and \ion{N}{1} abundances along these sight lines, and in
\S\ 6 we re-examine the \ion{H}{1} column densities. Section 7 contains
a discussion of the results and conclusions.

\section{The Data}

\subsection{\fuse\ Observations}

Information for the \fuse\ observations is given in Table 2.  The
\fuse\ mission and instrument are described by \cite{m00} and
\cite{s00}. The spectra were obtained through the large
(30\arcs$\times$30\arcs) LWRS aperture. \fuse\ consists of 4
co-aligned optical channels, two optimized for longer wavelength (LiF1
and LiF2; 1000 -- 1187~\AA) and two optimized for shorter wavelengths
(SiC1 and SiC2; 905 -- 1100~\AA). All of the deuterium lines used in
this paper are located in the SiC channels. The Ly$\beta$ \ion{D}{1}
line was not detected in either spectrum because it was blended with
the strong \ion{H}{1} line, but some lines of other elements used in
our analysis are located in the LiF channels.  Between exposures the
focal plane assemblies were moved so that the spectra sampled
different portions of the detector. This shifting lessens the effects
of detector defects when the individual spectra are combined.

The raw spectra were processed through the \fuse\ calibration pipeline
(CALFUSE v2.1.6) at the Johns Hopkins University.  The pipeline
screens data for passage through the South Atlantic Anomaly and low
Earth limb angle pointings and corrects for thermal drift of the
gratings, thermally-induced changes in the detector read-out
circuitry, and Doppler shifts due to the orbital motion of the
satellite. Finally, the pipeline subtracts a constant detector
background and applies wavelength and flux calibration. After the
pipeline reduction the individual spectra were co-added to produce the
final calibrated spectrum. The data from the four channels were
analyzed separately, because there are slight differences in the
spectral resolution between channels, which can cause problems if the
channels are co-added. This has the benefit of allowing at least two
measurements of each line, a further safeguard against detector
defects.  

The presence of bright airglow lines at specific regions of the
spectrum has resulted in a long-term depletion of electrons in the
detector in those regions, and this can cause localized shifts in the
spectrum in the dispersion direction. This effect, called ``x-walk,''
may change the shapes and areas of absorption lines. The effect is
more pronounced in certain \fuse\ detectors. We have avoided using the
equivalent widths of absorption lines in regions where the x-walk
effect is pronounced, so these detector artifacts do not affect the
analysis in this paper. The x-walk distortion is described in the {\em
FUSE Observer's Guide} available at the \fuse\ website
(\url{http://fuse.pha.jhu.edu/}).

Figure 1 shows a portion of the \fuse\ spectrum of HD~195965, and
Figure 2 shows the same spectral region for HD~191877. At
$\lambda \sim920$~\AA, the shortest wavelengths used in this paper, the signal
to noise ratio is $\ga$30:1 per $\sim$20 \kms\ resolution element in
both spectra, and is considerably higher at longer wavelengths.

\subsection{STIS Observations}

We observed HD~195965 with the Space Telescope Imaging Spectrograph (STIS)
on the \hst\ on 2001~October~9 as part of observing program GTO-8487 
(P.I.: Moos).  Brief information about the observation is given in Table~2.  
The spectrum was obtained in ACCUM mode using the FUV MAMA, the E140H grating 
($\lambda_c = 1307$\,\AA), and the $0.1\arcsec\times0.03\arcsec$ aperture.  
We reduced the data using CALSTIS (v2.12a) with all of the normal 
reduction procedures enabled in the standard data pipeline, including the 
two-dimensional scattered light removal algorithms (see Landsman \& Bowers 
1997; Lindler \& Bowers 2000; see also Howk \& Sembach 2000).  A description 
of the STIS pipeline processing steps can be found in the $HST$ Data Handbook \citep{b02}.

The extracted STIS spectrum has $S/N \sim 25-30$ per spectral resolution 
element.  The spectral resolution of $\sim2.6$ \kms\ 
(R = $\lambda/\Delta\lambda \sim 200,000$) is $\approx11$ times better than 
that available with $FUSE$.  The absolute wavelength scale is accurate to 
$0.5-1.0$ pixel, or about 3-6 m\AA\ ($\sim 0.6-1.3$ \kms; see 
Leitherer et al. 2001). The post-processed STIS data have very low 
residual backgrounds.  We find that the strongly saturated cores of the 
\ion{O}{1} 1302.169 \AA\ and \ion{C}{2} 1334.532 \AA\ lines have 
residual intensities of $\lesssim 1$\%. 

\subsection{Ground-based Observations}

High-resolution echelle spectra of HD~195965 and HD~191877 were taken
with the 2.7~m telescope of the University of Texas McDonald
Observatory on 2001 December 29.  The ``2dcoud\'{e}" spectrograph
\citep{t95} was used with 1 camera setup and a Tektronix charge-coupled
device (CCD) as the detector, which provided high spectral resolution
(R = $\lambda/\Delta\lambda\approx$ 160,000; $\Delta$v $\approx$ 2 km
s$^{-1}$) and covered the wavelength range 6000~\AA\ to 8600~\AA.
This high resolution mode yielded spectra containing 12 disjoint
echelle orders approximately 40~\AA\ wide, separated by approximately
150~\AA\ at 6000~\AA.  The inter-order separation increased with
wavelength until there was approximately 280~\AA\ between orders at
near infrared wavelengths.  This setup provided spectra on
interstellar \ion{K}{1}.  The resolution was determined from the full
width at half maximum (FWHM) of Thorium-Argon (Th-Ar) calibration lamp
lines.

The data were reduced in a standard way utilizing the NOAO SUN/IRAF
software (v2.11.3).  Dark, bias, and flat field lamp exposures were
taken to remove any instrumental effects due to the CCD
detector.  Comparison spectra were taken periodically throughout the
night, typically every two hours.  The average bias exposure was
subtracted from all raw stellar, comparison (Th-Ar) and flat images.
The scattered light was fitted by low order polynomials in both the
dispersion and spatial directions, and removed.  Pixel to pixel
sensitivity variations were removed by dividing the normalized average
flat field into the stellar spectra. The extracted spectra were placed
on a heliocentric wavelength scale using the Th-Ar comparison spectra
and corrections for the Earth's motion.  The spectra were co-added and
normalized to unity yielding a final spectrum with high signal to
noise, approximately 100:1. Figure 3 shows the spectral region near
the \ion{K}{1} 7698.964~\AA\ line for both sight lines after
correcting the velocities to the Local Standard of Rest (for HD~195965
$V_{LSR} = V_{helio} + 16.7$ \kms, and for HD~191877 $V_{LSR} =
V_{helio} + 17.4$ \kms).

\section{Properties of the Sight Lines}

HD~195965 ($l$=85.71$^\circ$, $b$=5.00$^\circ$) was classified by
Morgan, Code, \& Whitford (1955) as type B0V. It is listed by
\cite{hum78} as belonging to the Cygnus OB7 association, which is at a
distance of $\sim$830 pc in the Local spiral arm. \cite{ds94} give a
distance of 794~pc from spectroscopic parallax (with an uncertainty of
$\sim25\%$). At this distance it would be 69 pc above the Galactic
plane, still well within the thin disk, where most of the cool gas is
located. The Hipparcos catalog lists an uncertain distance of the star
at d$=520\pm^{230}_{120}$ pc \citep{esa97}. \cite{ds94} find log
$N$(HI)=20.9 based on \iue\ data.

HD~191877 ($l$=61.57$^\circ$, $b$=-6.45$^\circ$) is type B1.0Ib
\citep{mcw55}, and spectroscopic parallax gives a distance of
$\sim$2200~pc \citep{ds94}, again with an uncertainty of
$\sim25\%$. This places the star on the edge of the Sagittarius arm
and 250 pc below the plane of the Milky Way, well out of the thin disk
(scale height $\sim100$ pc, Dickey \& Lockman 1990). It is likely that
most of the gas along this sight line lies within $\sim1000$ pc of the
Sun. This is supported by the fact that the \ion{H}{1} column toward
HD~191877 found by \cite{ds94} (log $N$(HI)=20.9) is similar to that
toward HD~195965, even though the distance to HD~191877 is more than
twice as great as the distance to HD~195965. The two sight lines
therefore likely probe a similar distance range in the disk. The
Hipparcos distance to HD~191877 is d$>1160$ pc \citep{esa97}.

The ground-based optical spectra of HD~195965 and HD~191877 in Figure
3 reveal that there are multiple clouds along the line of sight to
each star. Along both sight lines there appear to be two major
components of \ion{K}{1}, as well as several weaker components. The
STIS spectrum of HD~195965 (Figure 4) confirms the presence of
multiple velocity components. Two components are visible in the
\ion{O}{1} 1355.598~\AA\ profiles, and the \ion{S}{1} 1425.030~\AA\
profile shows that the strongest of these is actually two
components. The structure in the \ion{S}{1} line matches well the
\ion{K}{1} component structure. The \ion{Cl}{1} line shows two additional
components at more negative velocities, only one of which is strong
enough to appear in the \ion{O}{1} 1355.598~\AA\ line. The \ion{Cl}{1}
line is usually a good tracer of H$_2$ \citep{j74}, so it appears that
the molecular hydrogen velocity structure along the sight line is not
complex. The strong \ion{O}{1} 1302.168~\AA\ line is completely
saturated and shows absorption extending over $\pm$30~\kms, along with
damping wings at even higher velocities.

Both sight lines contain a significant amount of H$_2$. Most of the
unlabeled features in Figures 1 and 2 are H$_2$ absorption
lines. Lines from the lower rotational levels ($J=0$,1,2) are mostly
saturated. These contain most of the H$_2$ column density, so we did
not attempt to determine the H$_2$ abundance. However, based on the
\ion{Cl}{1} column toward HD~195965, we estimate roughly
$(5-10)\times10^{19}$~cm$^{-2}$ on that sight line \citep{j74}. The
HD~191877 sight line probably contains a similar amount. There are
clearly multiple components along the HD~195965 sight line (revealed
by the \ion{Cl}{1} line), but they fall within $\pm$15~\kms, so at
\fuse\ resolution the H$_2$ distribution can be treated as a single
component (see \S\ 4.1).

The widths of the \ion{K}{1} lines in the spectra of both stars (and
the STIS lines in the HD~195965 spectrum) are larger than the
resolution of the data. This suggests that there are no strong, narrow
(unresolved) features on either sight line. We can use the STIS data
for HD~195965 to show this more conclusively for that sight
line. Figure 5 shows the apparent column density of \ion{S}{1} derived
from two different lines: 1295.653~\AA\ and 1296.174~\AA. Although the
oscillator strengths differ by a factor of $\sim$3 \citep{m91}, the
derived column density is very similar for the two lines, supporting
the conclusion that narrow, saturated features are not present toward
HD~195965. (See Savage \& Sembach 1991 for a discussion of apparent
column density profile comparisons.) In other words, while cold clouds
are clearly present, they do not contain a significant amount of the
material along these sight lines. Most of the column arises in the
diffuse ISM.

To test the effects of the lower resolution of \fuse\ on the analysis
of blended lines, we smoothed the absorption profiles of the
\ion{S}{1} 1295.653~\AA\ and 1296.174~\AA\ lines in the STIS data by
15 \kms\ to approximate FUSE resolution. We then determined the column
densities by summing the apparent optical depth over the line
profile. The total column density derived from the 1295.653~\AA\ line
is $N$(\ion{S}{1})=$(1.21\pm0.02)\times10^{13}$ cm$^{-2}$, while that
derived from the 1296.174~\AA\ line is
$N$(\ion{S}{1})=$(1.34\pm0.05)\times10^{13}$ cm$^{-2}$. Both values
are well within the $1\sigma$ uncertainties of the column densities
derived from the unsmoothed profiles,
$N$(\ion{S}{1})=$(1.26\pm0.08)\times10^{13}$ cm$^{-2}$ from the
1295.653~\AA\ line, and $N$(\ion{S}{1})=$(1.33\pm0.18)\times10^{13}$
cm$^{-2}$ from the 1296.174~\AA\ line. The measured equivalent widths
are $17.2\pm0.9$~m\AA\ and $7.4\pm1.0$~m\AA, similar to the weaker
\ion{N}{1} and \ion{O}{1} lines along the HD~195965 sight line.  We
also carried out this test on the \ion{K}{1} lines shown in Figure
3. Toward HD~191877, the column densities from the unsmoothed profile,
$(4.02\pm0.06)\times10^{11}$ cm$^{-2}$, agrees well with the column
density derived from the smoothed profile,
$(3.97\pm0.09)\times10^{11}$ cm$^{-2}$. The test was harder to perform
on the HD~195965 sight line because of residuals from poorly removed
airglow lines, but the column density derived from the unsmoothed
profile, $(3.24\pm0.16)\times10^{11}$ cm$^{-2}$, is in marginal
agreement with that derived from the smoothed profile,
$(2.91\pm0.19)\times10^{11}$ cm$^{-2}$. With measured equivalent
widths of $49.7\pm2.1$~m\AA\ and $62.0\pm1.0$~m\AA\ toward HD~195965
and HD~191877, respectively, these lines are similar in strength to
the stronger \ion{O}{1}, \ion{N}{1}, and \ion{D}{1} lines along both
sight lines. These tests provide more evidence that there are no
unresolved, saturated components along either sight line.

\section{The D~I Column Density}

The simplest approach to analyzing the chemical composition along a
line of sight is to measure the equivalent widths of all lines of a
species, assume the material lies in a single component, and construct
a Doppler-broadened curve of growth (COG). Since the \fuse\ bandpass
contains all of the transitions of \ion{D}{1} except for Ly$\alpha$,
in principle this is possible using the current data. In reality, many
of the \ion{D}{1} transitions are not usable due to interference from
neighboring absorption lines. Table 3 lists some of the important
interfering lines for the unusable \ion{D}{1} transitions along these
sight lines. The lower order \ion{D}{1} lines are blended with strong
\ion{H}{1} lines, while higher order lines are often blended with
strong H$_2$ lines.  However, in the spectra of each sight line there
are four \ion{D}{1} transitions that were either free of blending or
were blended with weak H$_2$ lines that could be removed:
$\lambda$919.102, $\lambda$920.713, $\lambda$922.900, and
$\lambda$925.974. A more complex but potentially more accurate method
is to model the absorption lines with a profile fitting program. This
technique can be especially advantageous when the velocity structure
of the cloud components is complicated. In this section we describe
both of these approaches to determining the
\ion{D}{1} abundances toward HD~195965 and HD~191877.

\subsection{Molecular Hydrogen Removal}

There are many H$_2$ lines in the spectral range around the deuterium
lines, and several can potentially affect the measurement of the D I
equivalent widths. The lines that affect the four \ion{D}{1} lines
used in this analysis are listed in Table 4. We include only lines
with rotational level J$\le$5, because no higher $J$ level lines were
found in the spectra. The lines listed in Table 4 are expected to be
strong enough to affect the \ion{D}{1} lines. The 920.803~\AA\ line
does not fall within the 920.713~\AA\ \ion{D}{1} line, but it could
affect the continuum on the red side of the line.

To remove the contamination, a line of the same rotational level in a
clean part of the spectrum was found to use as a template, following
\cite{j99}. The template line was chosen from the same detector
segment as the line being corrected, in order to avoid differences in
the line spread function between segments.  The parameters of the
template lines are listed in Table 4. A Gaussian function was fitted
to the template line, and the fit was scaled in optical depth space by
the ratio of the $f\lambda$ values to have the strength appropriate
for the contaminating line. The template line was chosen to have as
close to the same value of $f\lambda$ as the contaminating line as
possible in order to avoid saturation effects. The scaled fit was then
shifted to the wavelength of the contaminating line and divided out of
the observed spectrum around the D I lines. Another H$_2$ line close
in wavelength to the \ion{D}{1} line was used to register the position
of the scaled line before dividing it out, in order to compensate for
any relative errors in the wavelength scale. Figures 6 and 7 show the
corrected profiles (solid lines) along with the uncorrected profiles
(dashed lines) for both sight lines as a function of LSR velocity.

\subsection{Curve of Growth Analysis}

Once the H$_2$ lines were removed, it was necessary to determine
the location of the continuum at the position of the \ion{D}{1}
absorption lines. This was done by fitting low order polynomials
across the lines. The situation is complicated by the fact that the
\ion{D}{1} lines fall on the blue damping wings of the associated
\ion{H}{1} lines. We used the steeply-sloped blue sides of the
\ion{H}{1} lines to constrain the fits, and when possible used the red
wings of the \ion{H}{1} lines as a guide in determining the optimal
continuum. Figures 6 and 7 show the absorption profiles of the four
\ion{D}{1} lines after correction for H$_2$ for HD~195965 and
HD~191877, respectively, along with the adopted continuum
placement. 

We measured the equivalent widths of the \ion{D}{1} lines using the
procedures described by \cite{ss92}. The lines were integrated over a
velocity range that depended upon the details of the local continuum
and line blending. The equivalent widths are listed in Table 5 for
HD~195965 and Table 6 for HD~191877. A theoretical single-component
COG was fit to the equivalent widths by minimizing $\chi^2$ while
varying the column density $N$ and the Doppler parameter $b$. The
final COGs are shown in Figures 8 and 9 for HD~195965 and HD~191877,
respectively. The final uncertainties in the equivalent widths were
determined from the dispersion of the points about the COG, assuming
that a single component COG is appropriate for the sight lines. These
error bars are a combination of statistical uncertainties and
continuum fitting errors, as described by \cite{ss92}, and include the
effects of fixed-pattern noise.

Placing the continuum near the 925.974~\AA\ line was problematic in
the spectra of both stars, because of absorption from other elements
on the blue side of the absorption line (see Figures 6 and 7). This
results in large uncertainties in the equivalent widths. Because of
this the 925.974~\AA\ line is not given as much weight as the other
lines in the COG fitting. Table 7 contains the COG \ion{D}{1} column
density results for both sight lines.

The inset in each COG figure shows a contour plot of the
$\Delta\chi^2$ values in the $N$ and $b$ dimensions, with contours of
the minimum $\Delta\chi^2$+1, 4, and 9 shown (corresponding to 1, 2,
and 3$\sigma$). The quoted error bars for $N$ shown in Table 7 are the
maximum extent of the projection of the 2$\sigma$ contour onto the $N$
axis, and likewise for $b$ (listed in the captions to Figures 8 and
9).

\subsection{Absorption Profile Fitting}

Column densities were also measured independently with the profile
fitting procedure Owens.f, following the method described by
\cite{h02}.  Briefly, the software simultaneously fits Voigt profiles
to all of the unsaturated interstellar lines detected in all \fuse\
segments, using a $\chi^2$ minimization procedure with many free
parameters. The free parameters characterize the physical properties
of the interstellar clouds (radial velocities, column densities,
temperatures, micro-turbulent velocities) as well as the continuum
shapes (modeled with polynomials) and possible instrumental effects
(line spread function widths, wavelength solution variations). The
error bars are obtained using the standard $\Delta\chi^2$ method; they
include statistical as well as systematic uncertainties \citep{h02}.

The fits include the \ion{D}{1} lines 916.6~\AA, 919.1~\AA, and
920.7~\AA, the \ion{N}{1} lines 951.1~\AA, 951.3~\AA, and 959.5~\AA,
as well as \ion{Fe}{2}, and molecular H$_2$, HD, and CO lines. No
\ion{O}{1} lines were included in the fits because they are all
saturated. Since the H$_2$ lines were included in the fits, the lines
were {\it not} removed as for the COG analysis (\S\ 4.1).  Samples of
the fits are plotted on Figure 10, which presents two spectral windows
for each target. The final fits include 30 and 22 spectral windows for
HD195965 and HD191877, respectively.  The results for $N$(\ion{D}{1})
and $N$(\ion{N}{1}) are reported in Table~7.  We obtain $\log
N$(\ion{Fe}{2})$~=14.81\pm0.02$ and $14.95\pm0.04$ ($2\sigma$) for
HD~195965 and HD~191877, respectively.

\section{The O~I and N~I Column Densities}

We measured the equivalent widths of \ion{O}{1} and \ion{N}{1} lines
in the \fuse\ spectra using the same techniques as described for
\ion{D}{1}. The measured equivalent widths are listed in Table 5 for
HD~195965 and Table 6 for HD~191877.  Figure 11 shows single-component
COGs for \ion{O}{1} and \ion{N}{1} toward HD~195965. The \ion{O}{1}
lines in the \fuse\ bandpass that are not affected by H$_2$ absorption
do not cover a large range of oscillator strengths, and they mostly
fall on the flat part of the COG. The STIS spectrum of HD~195965 adds
two \ion{O}{1} lines: the 1355.598~\AA\ line on the linear part of the
COG, and the 1302.169~\AA\ line, which exhibits damping wings and is
on the square root portion of the COG. The STIS lines cover a larger
range of oscillator strengths, so the \ion{O}{1} abundance toward
HD~195965 is well constrained by the \fuse+STIS COG. The \ion{N}{1}
lines in the \fuse\ spectra also cover a large range of oscillator
strengths, so an accurate determination of the column density is
possible. The derived column densities for \ion{O}{1} and \ion{N}{1}
toward HD~195965 are listed in Table 7.

Figure 12 shows the COGs for \ion{O}{1} and \ion{N}{1} toward
HD~191877. As was the case for HD~195965, the \ion{N}{1} abundance is
well constrained, but the \ion{O}{1} abundance toward HD~191877 is
less certain because of the lack of STIS data. A single-component COG
fit to the \fuse\ points yields a column density of
log~$N$(\ion{O}{1})$=17.24\pm^{0.42}_{0.22}$~cm$^{-2}$ and the Doppler
broadening parameter $b=8.2\pm^{0.5}_{0.9}$~\kms ($2\sigma$). We also
set a lower limit of 230~m\AA\ ($3\sigma$) on the 1302.169~\AA\
equivalent width using an \iue\ spectrum of HD~191877 (see \S~6). This
point rules out the COG found for the \fuse\ points alone, and
restricts the \ion{O}{1} column density to
log~$N$(\ion{O}{1})$\ge17.42$~cm$^{-2}$, with $b=7.85$~\kms.  The
$b$-values derived from the COGs for \ion{N}{1} and \ion{O}{1} toward
HD~195965 are almost identical, suggesting that the velocity behavior
of these two species is coupled.  If we assume that this is also true
for the HD~191877 sight line and force $b=7.5$~\kms\ for the \ion{O}{1} COG
(the solid line in the top panel of Figure 12), the derived column
density of \ion{O}{1} is
log~$N$(\ion{O}{1})=$17.54\pm^{0.20}_{0.12}$~cm$^{-2}$
($2\sigma$). Although the HD~195965 sight line suggests that
\ion{O}{1} and \ion{N}{1} are coupled, we cannot know with certainty
whether this is also true along the HD~191877 sight line. This
exercise illustrates the fact that the \ion{O}{1} column density
toward HD~191877 is not well constrained by the lines in the \fuse\
spectrum. We adopt the larger $N$(\ion{O}{1}) column density as our
preferred value, but note that it should be treated with caution.

We also used the damping wings of the \ion{O}{1} 1302.169 \AA\ line in
the STIS spectrum of HD~195965 to make an independent determination of
the \ion{O}{1} abundance toward the star.  We fit a single component
Voigt profile to the normalized \ion{O}{1} line using an oscillator
strength f = 0.04887 and a natural damping constant $\gamma =
5.75\times10^8$~s$^{-1}$ (Morton 1991).  The best fit model has a
central velocity $v_{helio} = +2.1$ \kms, b = $6.3\pm0.3$ \kms, and
$\log N$(\ion{O}{1}) = $17.80\pm^{0.05}_{0.09}$ ($2\sigma$).  This
column density is in excellent agreement with the COG result listed in
Table~7.  The single-component fit to the \ion{O}{1} 1302.169 \AA\
line is shown in Figure~13.  The fit shown in the figure reproduces
the observed absorption profile core and damping wings well, with the
minor exception of a slight underestimate in the absorption strength
between +21 and +40 \kms.  This additional absorption is likely due to
very low column density gas along the sight line that does not
contribute much to the total column density; a component with an
\ion{O}{1} column density of $\sim10^{13}$ cm$^{-2}$ could account for
this minor discrepancy.  Experimentation with additional components
does not change the best-fit result substantially.  Performing a fit
with the \ion{O}{1} column density divided among two components
separated by $\sim$ 12 \kms\ as suggested by the \ion{O}{1} 1355.598
\AA\ line shown in Figure~4 yields a best fit nearly indistinguishable
from the single-component fit shown in Figure~13.

The \ion{O}{1} 1302.169 \AA\ line is very strong, and can be used to
test for the presence of high-velocity components that might
contribute to the column density of \ion{H}{1} but be outside of
velocity integration range for other elements. The most obvious
candidate is the component between +21 and +40 \kms, which contributes
only $N$(\ion{O}{1})$\sim10^{13}$ cm$^{-2}$, corresponding to
$N$(\ion{H}{1})$\sim3\times10^{16}$ cm$^{-2}$ using the O/H ratio of
$3.43\times10^{-4}$ from \cite{meyer01}, or $\sim3\times10^{-5}$ of the
total \ion{H}{1} column density (see \S\ 6). No other components this
strong are visible in Figure~13, so we conclude that high-velocity
\ion{H}{1} is not an issue for the HD~195965 sight line.

The STIS spectrum of HD~195965 contains both the optically thin
1355.598~\AA\ line and the damped 1302.169~\AA\ line. We can use this
fact to test for consistency between the oscillator strengths of the
two lines (see Sofia, Cardelli, \& Savage 1994). The \ion{O}{1} column
densities derived from these lines are
$N$(\ion{O}{1})$=(5.7\pm1.1)\times10^{17}$ cm$^{-2}$ and
$(6.3\pm0.5)\times10^{17}$ cm$^{-2}$, respectively (uncertainties are
$1\sigma$). These results are identical within the uncertainty, which
suggests that any error the oscillator strength of either line is less
than 20\% (the uncertainty in the 1355.598~\AA\ equivalent width).
The oscillator strength for the 1355.598~\AA\ line listed in Table 5
includes the correction from
\cite{wel99}.

\section{The H~I Column Density}

We have re-measured the \ion{H}{1} column densities for the HD~191877
and HD~195965 sight lines using existing STIS and \iue\ data.
Single-component Voigt profiles were fit to the \ion{H}{1} absorption
toward each star.  The fits were limited to the 1170 to 1260 \AA\
spectral region, with the overall continuum levels determined at
longer wavelengths, as needed.  For both sight lines, the fit is
insensitive to the assumption of a single absorbing component.  Adding
additional components or changing the $b$-value of the absorption
(assumed to be in the 1 to 100 \kms\ range) did not alter the fits
significantly since the quality of the fits is governed primarily by
the steep walls of the line cores and the broad radiation-damping
wings.  The Voigt profile fitting method has been described in detail
for \ion{H}{1} determinations in previous \fuse\ D/H papers (see Moos
et al. 2002, and references therein).  Our analyses follow many of
these same principles and methods.

For HD~195965, we used the STIS E140H data available in this
wavelength region (see \S2.2) to determine the \ion{H}{1} column
density.  We paid special attention to making sure that the flux
calibration mapping at the ends of the individual echelle orders was
consistent between adjoining orders.  Scattered light is not a
significant source of uncertainty in the final measurement since we
used the damping wings of the lines rather than the saturated portion
of the line core to determine the \ion{H}{1} column density.  We find
a best-fit value of log $N$(\ion{H}{1}) = 20.95$\pm$0.05 ($2\sigma$)
for the fit to the high quality STIS data shown in Figure 14.  This
value is slightly higher than, but within the errors of, the former
value of 20.90+/-0.09 ($1\sigma$) found by \cite{ds94} using the
"continuum reconstruction" technique on existing \iue\ data.  The
reconstruction method and profile fitting methods are complimentary to
each other.  The fit is shown in the bottom panel of Figure 14,
together with the fits appropriate for the $\pm2\sigma$ column density
ranges (dashed lines).  For this sight line, the higher quality STIS
data more tightly defines the extent of the \ion{H}{1} absorption than
the \iue\ data, and as can be seen in the top panel of Figure 14, the
resulting fit is consistent with the \iue\ small aperture data
analyzed previously.

The primary difficulty in fitting the STIS profile of HD~195965 arises
in the wavelength region between 1222 and 1225 \AA, where the fit
slightly overestimates the amount of absorption present.  We explored
other models that fit this region better (at the expense of fit
quality at other wavelengths) and found that in all cases the
resulting column density was within the listed uncertainties.  Therefore,
while we are uncertain of the exact cause of the discrepancy in this
wavelength region, we believe the resulting \ion{H}{1} column density
is a robust estimate. Note that this portion of the STIS spectrum
contains the border between two echelle orders, and that the anomaly
is not apparent in the \iue\ spectrum.

For HD~191877, we re-examined the \iue\ small aperture data (SWP02837)
available for this sight line.  A more recent exposure (SWP148425)
taken through the small aperture clearly has problems with the flux
levels as a result of poor centering; we did not use this low quality
spectrum.  We find log $N$(\ion{H}{1}) = 21.05$\pm$0.10 ($2\sigma$),
which again is slightly larger than, but consistent with, the previous
estimate of 20.90+/-0.10 ($1\sigma$) found by \cite{ds94}.  The fit is
shown in Figure 15, along with the $\pm2\sigma$ error bars.  As can be
seen from this figure, a fit with a smaller column, like the one
preferred by the previous investigators, is clearly below the true
column density (see the $-2\sigma$ [log $N = 20.95$] model in the
figure).  Diplas \& Savage note that there may be a small amount
($\sim0.01$ dex) of contamination by stellar absorption in the
\ion{H}{1} Ly$\alpha$ profile.  This contamination does not affect our
$N$(\ion{H}{1}) estimate significantly.

\section{Discussion}

The sight lines analyzed in this paper are the longest ones for which
the \ion{D}{1} column densities have been determined with
\fuse. As such, they present the opportunity to look for variations in
\ion{D}{1}/\ion{H}{1}, \ion{D}{1}/\ion{O}{1}, and \ion{D}{1}/\ion{N}{1}
beyond the local ISM sight lines presented by \cite{m02}. Table 8
contains column density ratios of various elements toward both stars,
along with the mean values for the local ISM from \cite{m02}. In this
section we discuss the behavior of these column density ratios beyond
the local ISM.

\subsection{D/H}

The deuterium abundances along seven sight lines have been analyzed with
\fuse\ \citep{m02,k02,f02,s02,l02,nl02,w02,h02}. Five of these had both
\ion{D}{1} and \ion{H}{1} measurements, and the results were
consistent with a constant value of D/H in the local ISM. Earlier
\copernicus\ observations showed evidence for variations in D/H at
larger distances (see Vidal-Madjar \& Gry 1984 for a review), evidence
which was strengthened by subsequent \imaps\ investigations
\citep{j99,stfjvw00}. The main obstacle to verifying and understanding
this variation is the paucity of suitable deuterium sight lines that
probe distances $\ge$250 pc.

Figure 16 shows the D/H ratios for HD~195965 and HD~191877 plotted as
a function of distance from the Sun, along with five of the seven
original \fuse\ sight lines with reliable \ion{H}{1} measurements, and
several others from \hst, \imaps, and \copernicus\ (see Moos et
al. 2002).  The D/H ratios toward both stars are lower than the
average value found for the original seven \fuse\ sight lines. This
contrasts with the lack of variation seen in the local ISM. The new
sight lines support the existence of variations in the D/H ratio at
distances larger than a few hundred pc, as also suggested by
\imaps\ and\copernicus. 

In the simplest model of chemical evolution, stellar processing
(astration) leads to a decline in the D/H ratio because D is easily
destroyed in stars. Thus, the D/H ratio is expected to vary among
regions of the Galaxy that have undergone different amounts of star
formation in the past, provided that the gas in these regions has not
mixed ({\it e.g.} Tosi et al. 1998; Chiappini et al. 2002). In a small
volume like the local ISM ($r=100$ pc), \cite{m02} showed that the
time between supernova $t_{SN}$ is roughly equal to the mixing time
scale (assumed to be the sound crossing time $t_{s}$), suggesting that
the gas in such a region is well-mixed. If a larger region is
considered, say 1000 pc in radius to approximate the scales probed by
HD~195965 and HD~191877, the crossing time increases by a factor of
10, while the time between supernovae decreases by a factor of 100, so
that $t_{SN} \ll t_{s}$. Supernovae are not distributed randomly over
this region but are spatially correlated in regions of star formation,
meaning that substantial inhomogeneities will arise between
star-forming regions and more quiescent regions. For these reasons it
is not surprising to see variations over the distances probed by
HD~195965 and HD~191877.

A more precise treatment of mixing in the ISM was carried out by
\cite{am02}. Using numerical simulations, they found that the mixing
time scales were a few times 10$^8$ years. Chemical inhomogeneities in
the ISM are long-lived, so variations in D/H are expected over 1000~pc
length scales. In fact, since the mixing time scales increase only
slightly with length scale in their models, it is unlikely that the
local ISM ({\it e.g.,} the original seven \fuse\ sight lines) would
have a constant D/H value if any sources of inhomogeneities ({\it
i.e.,} supernovae) have occurred within the past $\sim10^8$ years.

We have investigated several alternative explanations for the low D/H
values. One possibility is that some of the D column is in
high-velocity components which are outside of the velocity integration
range used to determine $N$(\ion{D}{1}) \citep{j99}. We have used the
\ion{O}{1} 1302.169~\AA\ line in the STIS spectrum of HD~195965 to
rule this out for that sight line (see \S\ 5). Note that $\sim50\%$ of
the \ion{D}{1} column would have to be missed if the D/H ratio for the
HD~191877 sight line were as high as the \cite{m02} mean local ISM
value. Figure 7 shows that this is not the case, although we cannot
rule out a much smaller systematic offset for HD~191877.

A second possibility is that a significant amount of D is locked up in
HD molecules. To test this idea, we have measured the equivalent
widths of several strong, isolated Lyman series HD lines in the \fuse\
spectra: (3-0)~R(0)~1066.271~\AA, (5-0)~R(0)~1042.847~\AA, and
(4-0)~R(0)~1054.288~\AA. The column densities derived from the
equivalent widths are $\sim2.4\times10^{14}$ cm$^{-2}$ for the
HD~195965 sight line, and $\sim3.0\times10^{14}$ cm$^{-2}$ for the
HD~191877 sight line. For both sight lines this is $\la3.5$\% of the
total D column density, much lower than the uncertainty in $N$(\ion{D}{1}).

\subsection{O/H}

The D/O and O/H ratios toward HD~195965 in Table 8 are quite different
from the average values in the local ISM found by \cite{m02}. This is
a surprising result, since \cite{mjc98} found very little variation in
O/H in their survey, which covered distances and \ion{H}{1} column
densities similar to those of the sight lines in this paper.  Both the
\ion{O}{1} and \ion{H}{1} column densities toward HD~195965 were
determined through at least two independent methods, with very similar
results. The \ion{H}{1} was measured using both \iue\ and STIS data,
and the \ion{O}{1} was measured with a single-component COG which
combined \fuse\ and STIS data, as well as independent estimates from
the optically thin 1355.598~\AA\ line and the damped 1302.169~\AA\
line in the STIS spectrum. Both of these column densities are
therefore quite firmly established, and we believe that it is very
unlikely that there is a substantial systematic error in either
quantity.

HD~195965 appears to lie on a very high metallicity sight line, with
supersolar O/H \citep{sm01}. This may be a region where the signature
of stellar processed gas is still visible (as is not the case for the
local bubble). \cite{hum78} found that this star is in the Cygnus OB7
association. If so, it may be in a peculiar region, perhaps in a
metal-enriched cloud of ejecta from the star itself or a
neighbor. Since gas with high metallicity is thought to have undergone
stellar processing, D/H in such gas is expected to be low. This
expected anticorrelation between O/H (or N/H) and D/H was not seen in
the first seven \fuse\ sight lines, although it could have been masked
by the small number of sight lines studied \citep{st02}. Figure 17
compares the O/H and D/H value for HD~195965 with those of the
original seven \fuse\ sight lines. Unfortunately HD~195965 is the only
point at high metallicity, so no firm conclusions can be
drawn. However, it is possible that this sight line shows evidence of
the expected anticorrelation in the ISM (see also Steigman 2002). Note
that an error in the \ion{H}{1} column density toward HD~195965 would
tend to move that point along a diagonal line with a positive slope,
and thus could not produce the expected trend of decreasing D/H with
rising O/H.

The HD~191877 sight line lacks STIS data, so we could not constrain
the \ion{O}{1} and \ion{H}{1} abundances with multiple estimates. For
these reasons we are less confident in the \ion{O}{1} column densities
found for the HD~191877 sight line. As discussed in \S\ 5, the derived
log $N$(\ion{O}{1}) is very uncertain, although our preferred value,
found by coupling the $b$-values of \ion{N}{1} and
\ion{O}{1}, is consistent with the \fuse\ local ISM ratio and the
\cite{mjc98} ratio. Unfortunately, with the existing information, we
cannot make any strong conclusions about O/H toward HD~191877. 

\subsection{N/H}

The N/H ratios toward HD~195965 and HD~191877 in Table 8 are higher
than the average values in the local ISM found by \cite{m02}. Figure
18 shows a plot of N/H versus D/H along both sight lines, compared to
the four \fuse\ sight lines for which N/H was measured.  The figure is
very suggestive of an anticorrelation between D/H and N/H.

Complicating the interpretation of this plot is the fact N/H
ratio is susceptible to ionization effects, especially at low column
densities which do not provide adequate shielded from ionizing
photons, such as the column densities along the local ISM sight lines
\citep{j00}. The dashed line in Figure 18 shows the ISM value of N/H
found by \cite{mcs97}. The sight lines used in that study had higher
\ion{H}{1} column densities than the \fuse\ local ISM sight lines, so the
\fuse\ sight lines are very likely underestimating N/H, because a
significant fraction of N is ionized. The apparent trend in Figure 18
may simply reflect the tendency for $N$(\ion{N}{2})/$N$(\ion{N}{1}) to
decrease with increasing $N$(\ion{H}{1}).

However, an anticorrelation between D and N abundances would be
expected if D is destroyed in stars while N is produced ({\it e.g.}
Steigman et al. 2002). The overabundance of heavy elements accompanied
by the underabundance of D seems to fit well within a simple chemical
evolution scenario. While ionization effects probably play a role in
the high N/H ratio on these sight lines, they may not be the only
factor. A comprehensive N/H survey using \fuse\ spectra is currently
underway, and may shed light on this issue.

Several sight lines do not follow the expected trend of low D/H
accompanied by high N/H or O/H. \cite{s02} found a low O/H ratio
toward BD+28$^{\circ}$4211 with \fuse, while their D/H ratio is close to
the Local ISM mean. \cite{j00} found a low D/H toward $\delta$ Orionis
using \imaps, but also a somewhat low N/H ratio. These deviations
suggest that the simple picture of increasing heavy elements abundances
with decreasing D/H is not complete.

\subsection{Future Prospects for D/H Measurements}

Measuring D/H values in the Milky Way becomes increasingly difficult
as $N$(\ion{H}{1}) increases, perhaps reaching a fundamental confusion
limit. The COGs in Figures 8 and 9 show that the high order \ion{D}{1}
Lyman transitions are close to saturation. At higher column densities,
these transitions will fall on the flat part of the COG and will not
give precise column density information. Even higher order transitions
at shorter wavelengths, where the \ion{D}{1} lines might not yet be
saturated at higher column density, are not useful because the
\ion{H}{1} transitions begin to blend together. These facts make it
unlikely that many Galactic disk sight lines more distant than those
studied here will be useful for D/H measurements. However, the sight
lines analyzed in this paper demonstrate that D/H properties over the
distance range from 500 to 2500 pc can be explored with \fuse, even if
these sight lines contain a significant amount of H$_2$. This first
glimpse at D/H in this distance range suggests that it will be very
useful for understanding mixing and chemical evolution in the ISM.

The chemical evolution model of \cite{crm02} predicts an increase in
D/H with galactocentric radius, accompanied by a decrease in N/H and
O/H, leading to a pronounced gradient in D/O and N/O. The gas in the
outer parts of the galaxy has undergone less stellar processing, and
in fact \cite{crm02} predict that the D/H ratios in the outer disk
will approach the primordial value. The sight lines studied here have
roughly the same galactocentric radii as the local ISM sight lines of
\cite{m02}, so we cannot test this model, but with \fuse\
observations of sight lines in the inner and outer Galactic disk this
may prove possible.

\acknowledgments

We would like to thank Ed Jenkins, Gary Steigman, Jeffrey Linsky,
Cristina Oliveira, and Paule Sonnentrucker for enlightening
discussions. This work is based on data obtained for the Guaranteed
Time Team by the NASA-CNES-CSA \fuse\ mission operated by the Johns
Hopkins University. Financial support to U. S. participants has been
provided by NASA contract NAS5-32985. French participants are
supported by CNES. This work was partly supported by NASA Guaranteed
Time Observer funding to the STIS Science Team under NASA contract
NAS5-30403 and is based upon observations obtained with the NASA/ESA
Hubble Space Telescope, which is operated by the Association of
Universities for Research in Astronomy, Inc., under NASA contract
NAS5-26555. This work has made use of the profile fitting procedure
Owens.f developed by M. Lemoine and the FUSE French Team. GH would
like to thank Sylvestre Lacour and Arielle Moullet. DK thanks the
staff at McDonald Observatory, especially David Doss.

{}

\clearpage

\begin{figure}
\figurenum{1}
\epsscale{0.7}
\plotone{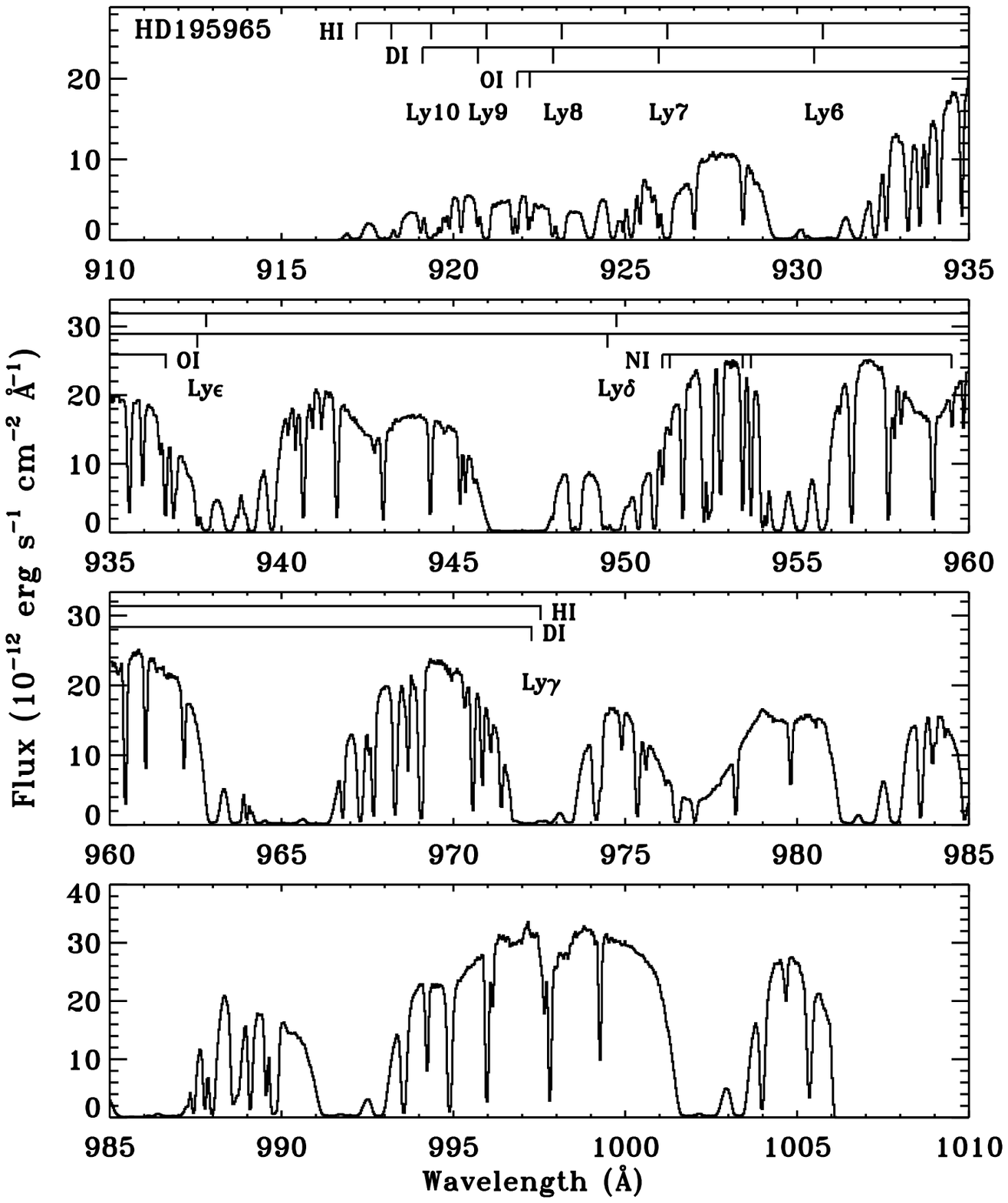}
\caption{A portion of the {\it FUSE} SiC2 spectrum of HD~195965. The Lyman transitions of H I and D I are labeled, as are the O I and N I transitions in this region of the spectrum that were used in the analysis. The D I transitions used in the curves of growth were Ly7, Ly8, Ly9, and Ly10. Most of the other features are H$_2$ absorption lines. The spectrum shown here is binned by 5 pixels to $\sim$10 km s$^{-1}$ bins for display.}
\end{figure} 

\begin{figure}
\figurenum{2}
\epsscale{0.7}
\plotone{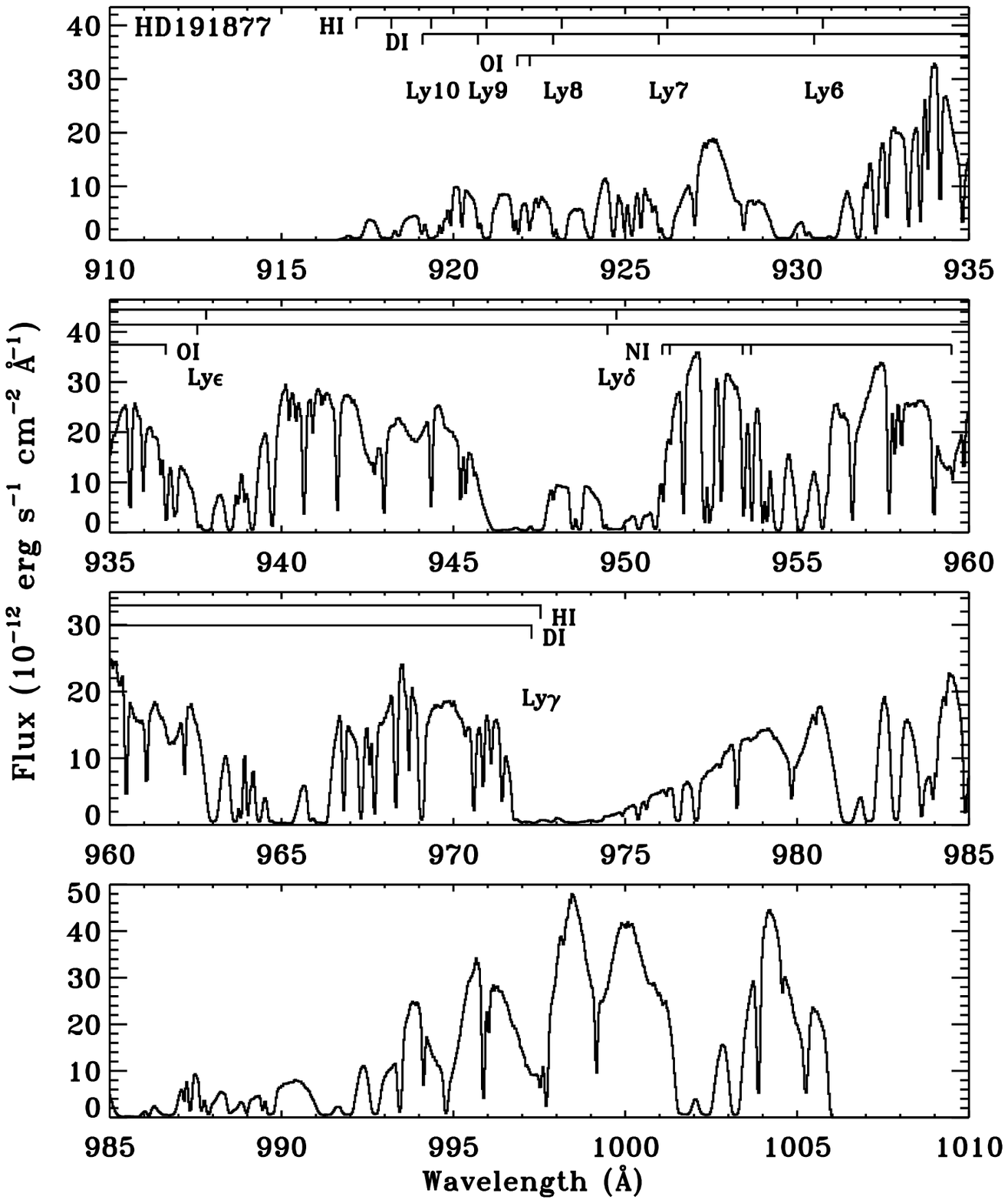}
\caption{A portion of the {\it FUSE} SiC2 spectrum of HD~191877. The Lyman transitions of H I and D I are labeled, as are the O I and N I transitions in this region of the spectrum that were used in the analysis. The D I transitions used in the curves of growth were Ly7, Ly8, Ly9, and Ly10. Most of the other features are H$_2$ absorption lines. The spectrum shown here is binned by 5 pixels to $\sim$10 km s$^{-1}$ bins for display.}
\end{figure} 

\begin{figure}
\figurenum{3}
\epsscale{0.70}
\plotone{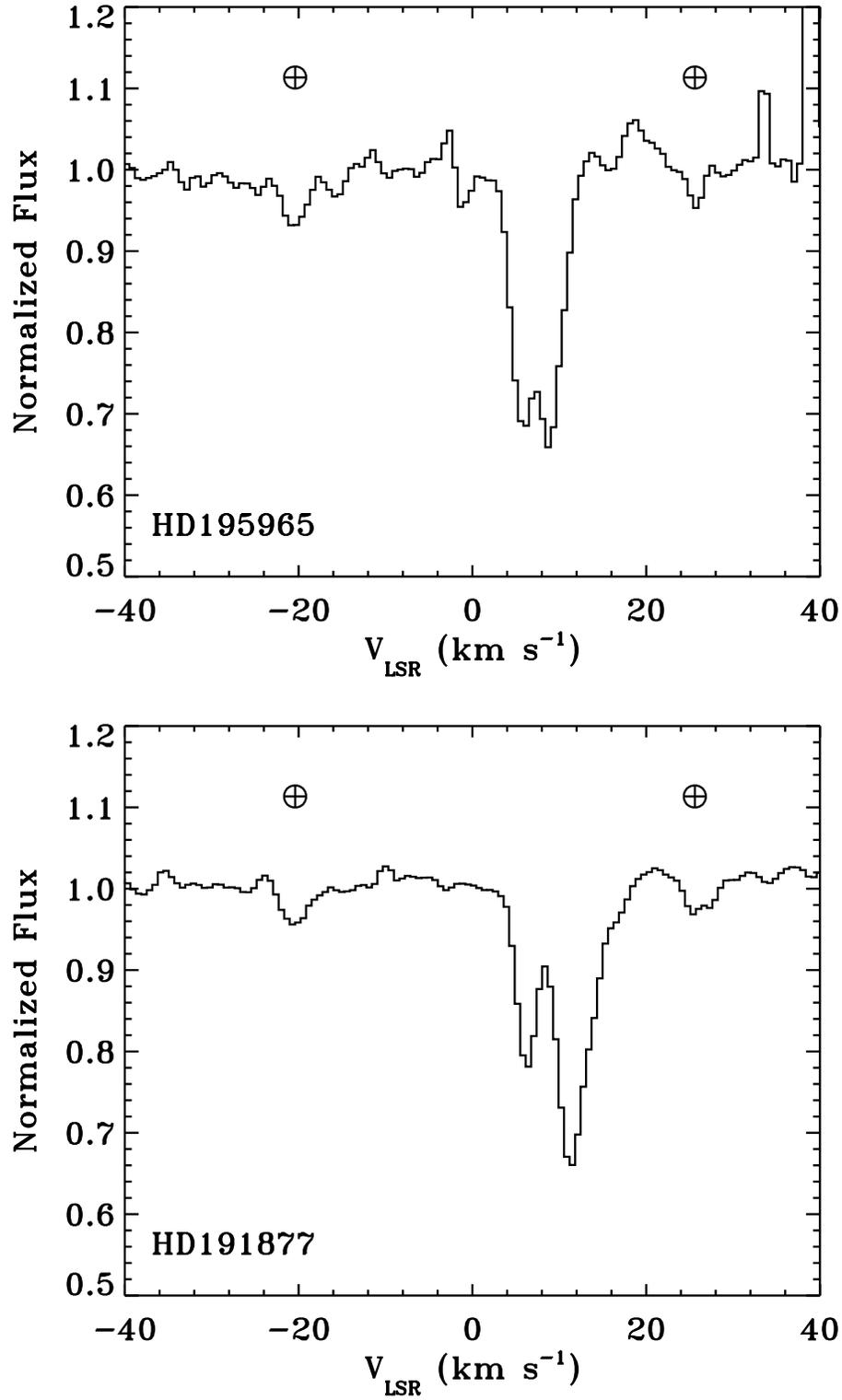}
\caption{Absorption profiles of the K~I 7698.964~\AA\ line for both stars. The absorption lines at -20 \kms\ and +28 \kms\ are remnants of atmospheric H$_2$O lines.}
\end{figure} 

\begin{figure}
\figurenum{4}
\epsscale{0.5}
\plotone{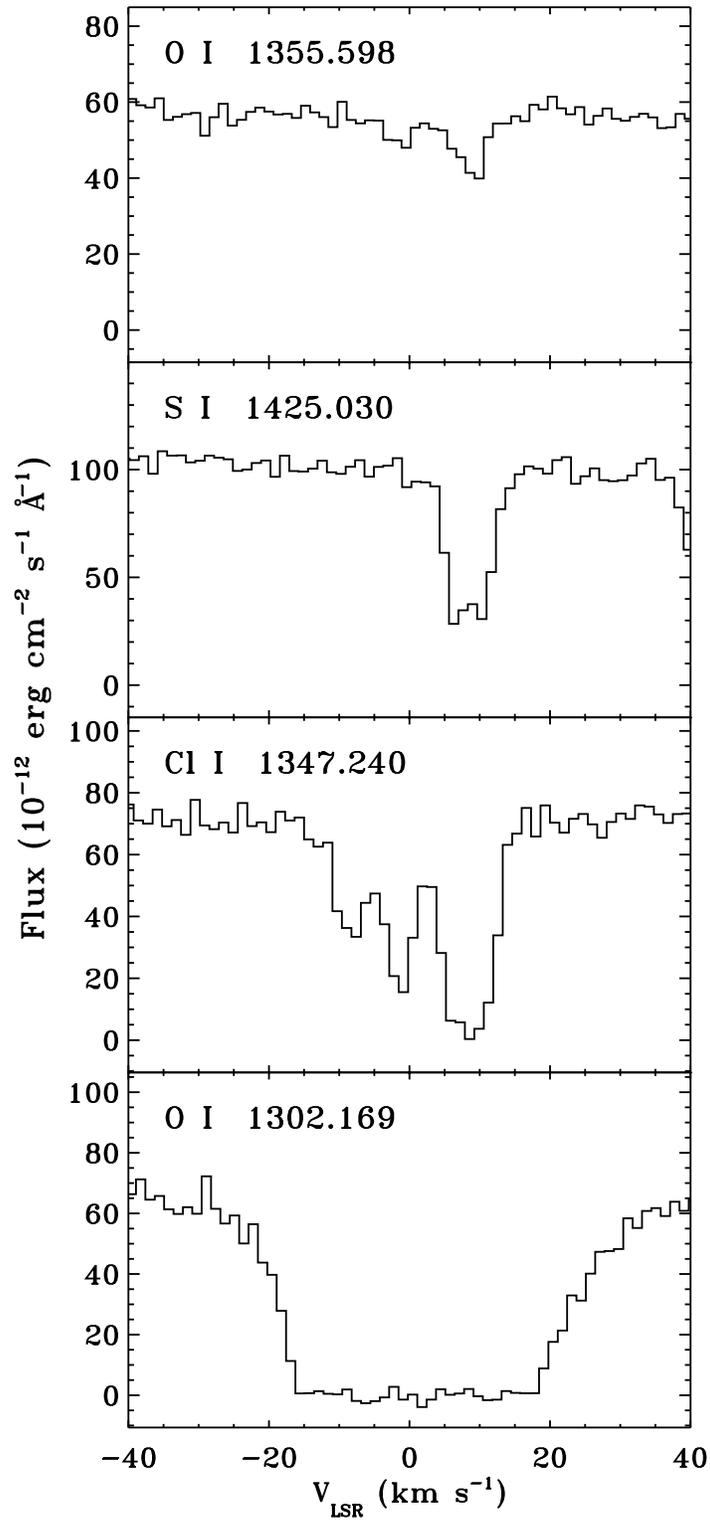}
\caption{Absorption profiles of selected lines from the STIS spectrum of HD~195965. }
\end{figure} 

\begin{figure}
\figurenum{5}
\epsscale{0.8}
\plotone{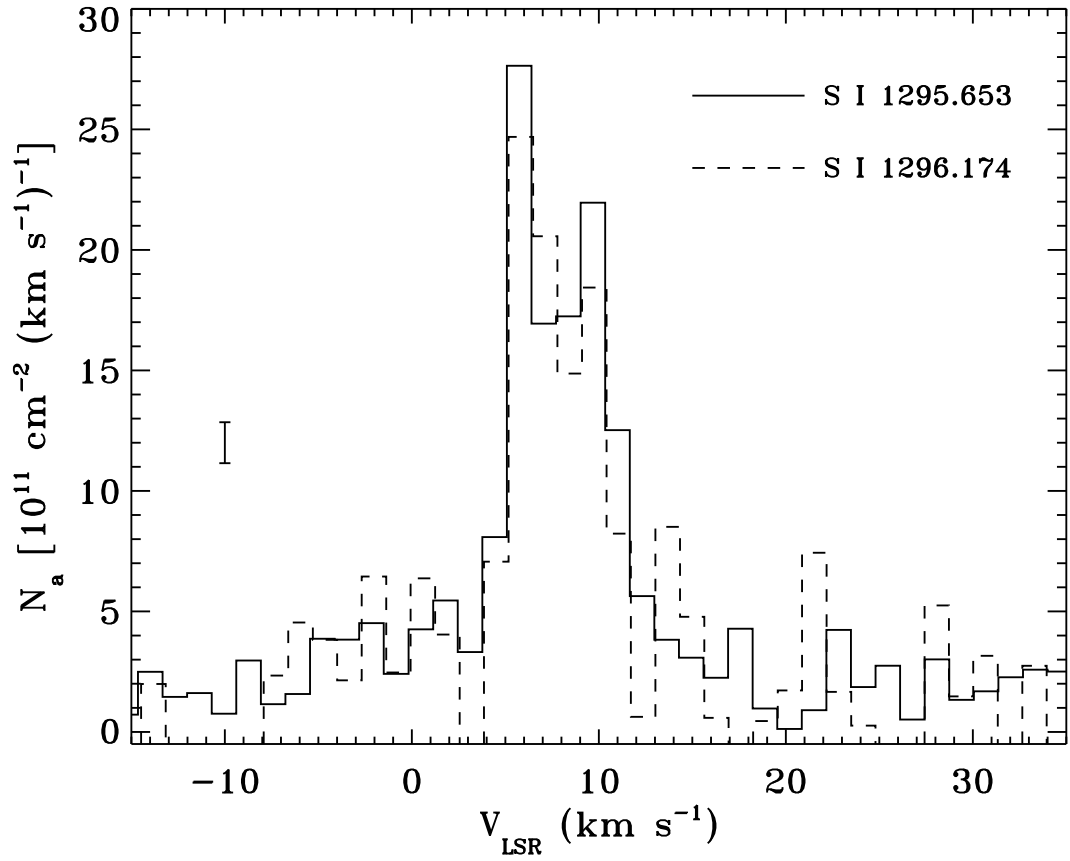}
\caption{Comparison of the apparent column density profiles derived from two S I lines in the STIS spectrum of HD~195965. The strengths ($f\lambda$) of the two lines differ by a factor $\sim$3, but the apparent column densities are similar at all velocities. The total column density between 3 and 13 \kms derived from the 1295.653~\AA\ line is $N$(S~I)=$(1.26\pm0.08)\times10^{13}$ cm$^{-2}$, while that derived from the 1296.174~\AA\ line is $N$(S~I)=$(1.33\pm0.18)\times10^{13}$ cm$^{-2}$. A representative error bar is shown.}
\end{figure} 

\begin{figure}
\figurenum{6}
\epsscale{0.7}
\plotone{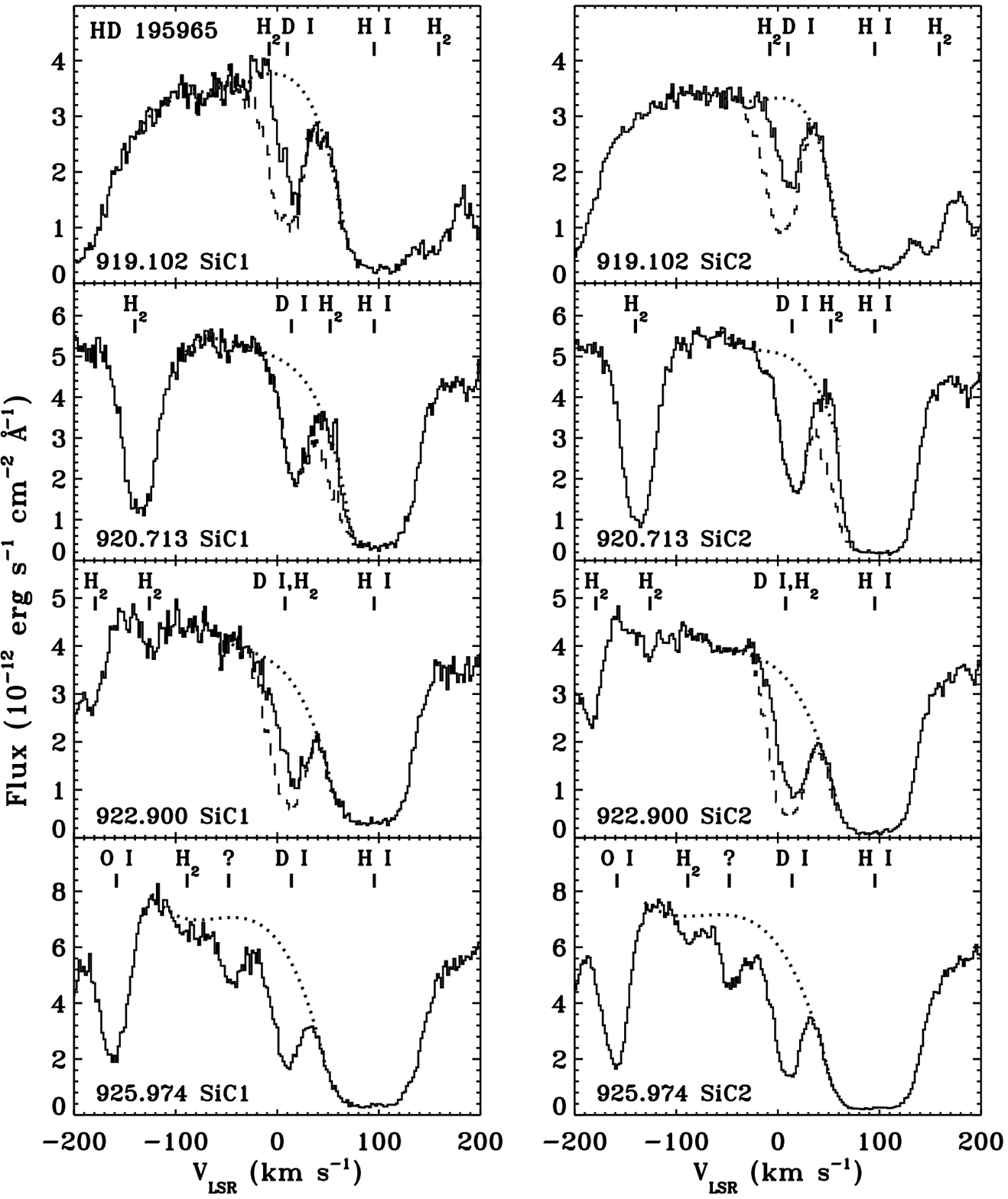}
\caption{Four D I absorption profiles in the {\it FUSE} HD~195965 spectra. The 919.102~\AA, 920.713~\AA, and 922.900~\AA\ profiles have been corrected for H$_2$ absorption. The uncorrected profiles are also shown (dashed lines). The 925.974~\AA\ profile is not affected by H$_2$ absorption. The adopted continua are shown as dotted lines. The feature at 925.780~\AA\ ($-45$~\kms\ in the bottom panels) is unidentified.}
\end{figure} 

\begin{figure}
\figurenum{7}
\epsscale{0.7}
\plotone{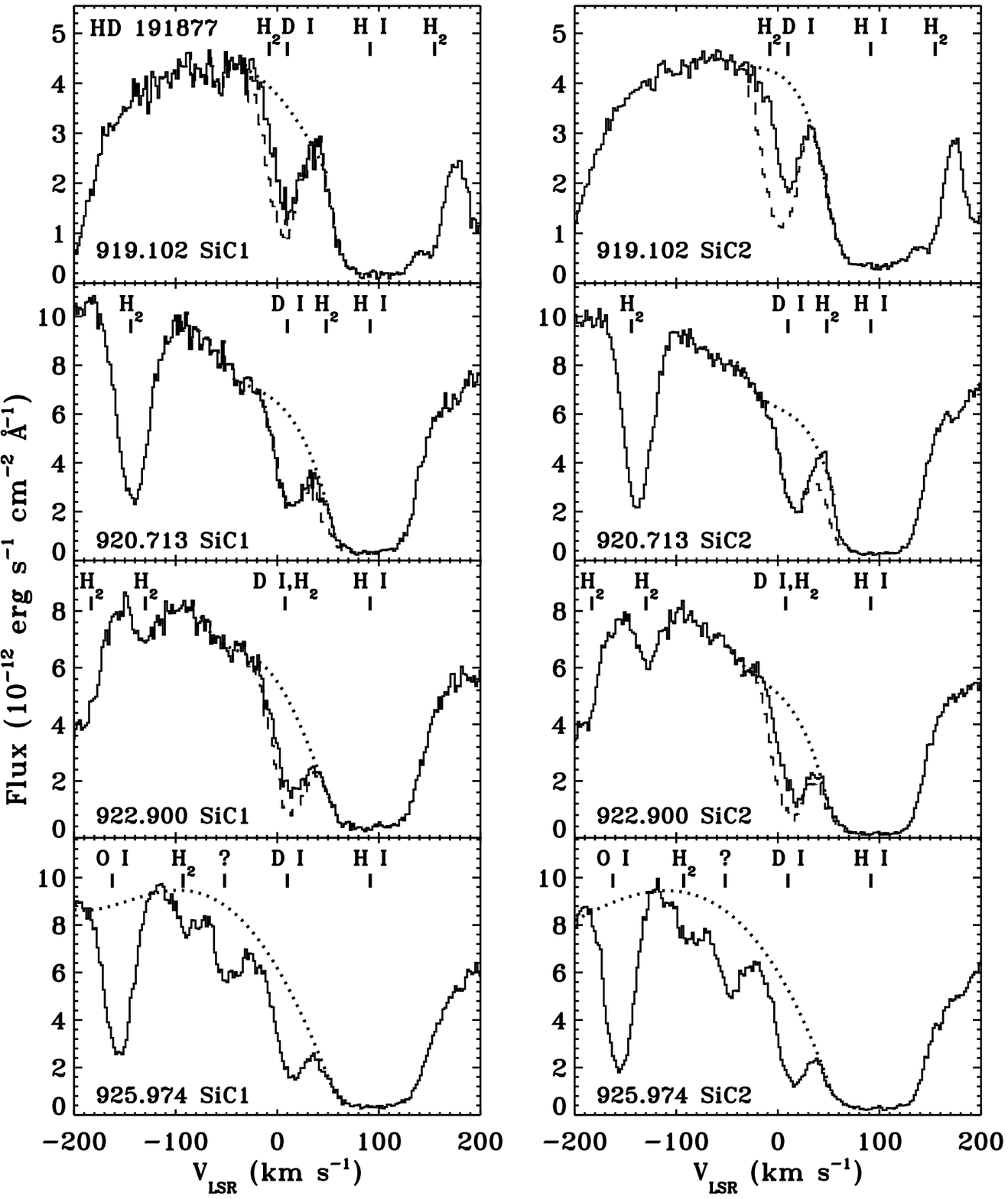}
\caption{Four D I absorption profiles in the {\it FUSE} HD~191877 spectra. The 919.102~\AA, 920.713~\AA, and 922.900~\AA\ profiles have been corrected for H$_2$ absorption. The uncorrected profiles are also shown (dashed lines). The 925.974~\AA\ profile is not affected by H$_2$ absorption. The adopted continua are shown as dotted lines. The feature at 925.780~\AA\ ($-45$~\kms\ in the bottom panels) is unidentified.}
\end{figure} 

\begin{figure}
\figurenum{8}
\epsscale{1.0}
\plotone{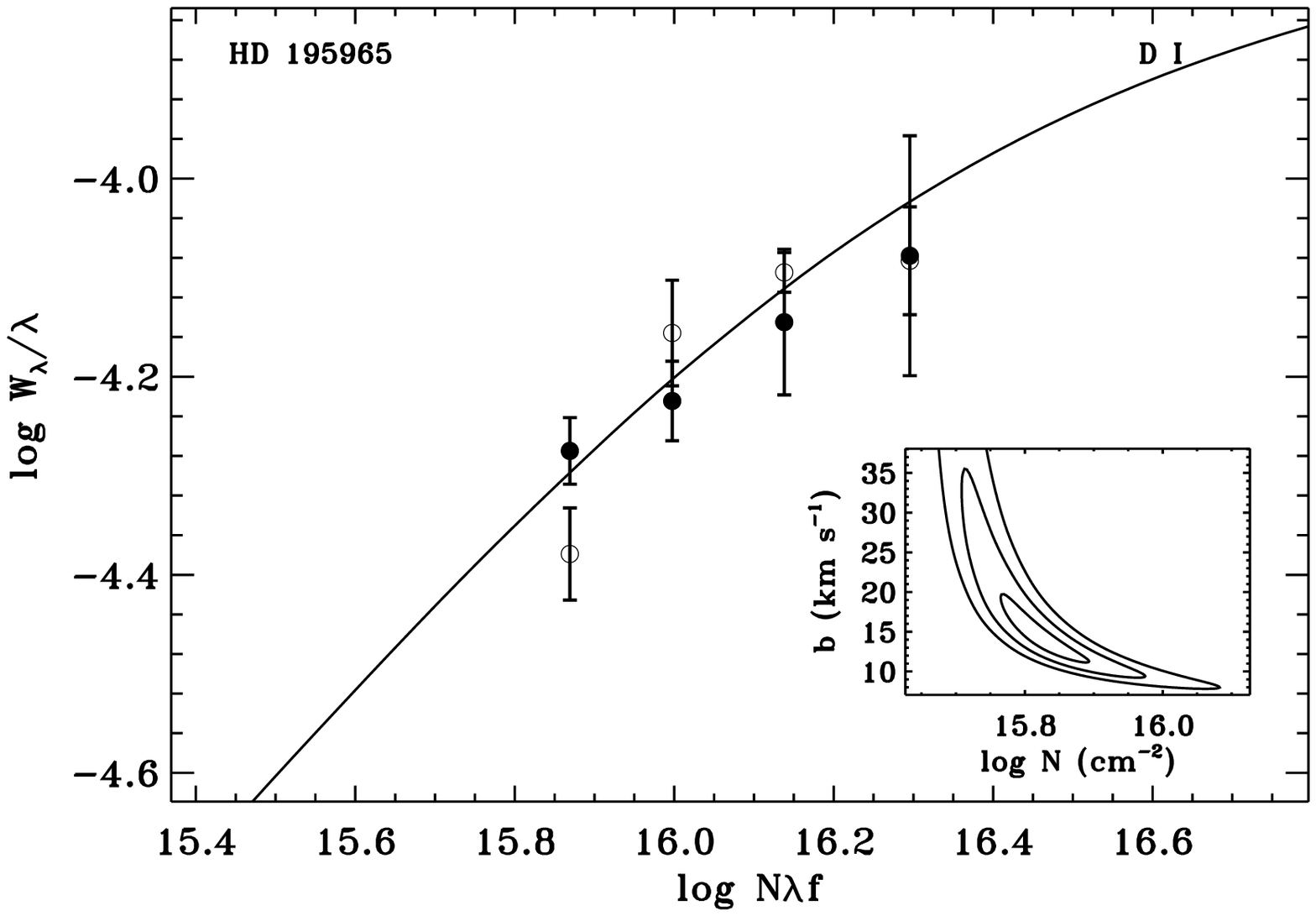}
\caption{The best fit single-component curve of growth for D I toward HD~195965, which gives log $N$(D I)=$15.83\pm^{0.14}_{0.11}$ (2$\sigma$) and  $b=14.1\pm^{21.4}_{4.7}$ km s$^{-1}$ (2$\sigma$). The filled circles are points measured from the SiC1 channel, and the open circles are points measured from the SiC2 channel. The inset shows a contour plot of the $\chi^2$ distribution, with the 1$\sigma$, 2$\sigma$, and 3$\sigma$ contours shown.}
\end{figure} 

\begin{figure}
\figurenum{9}
\epsscale{1.0}
\plotone{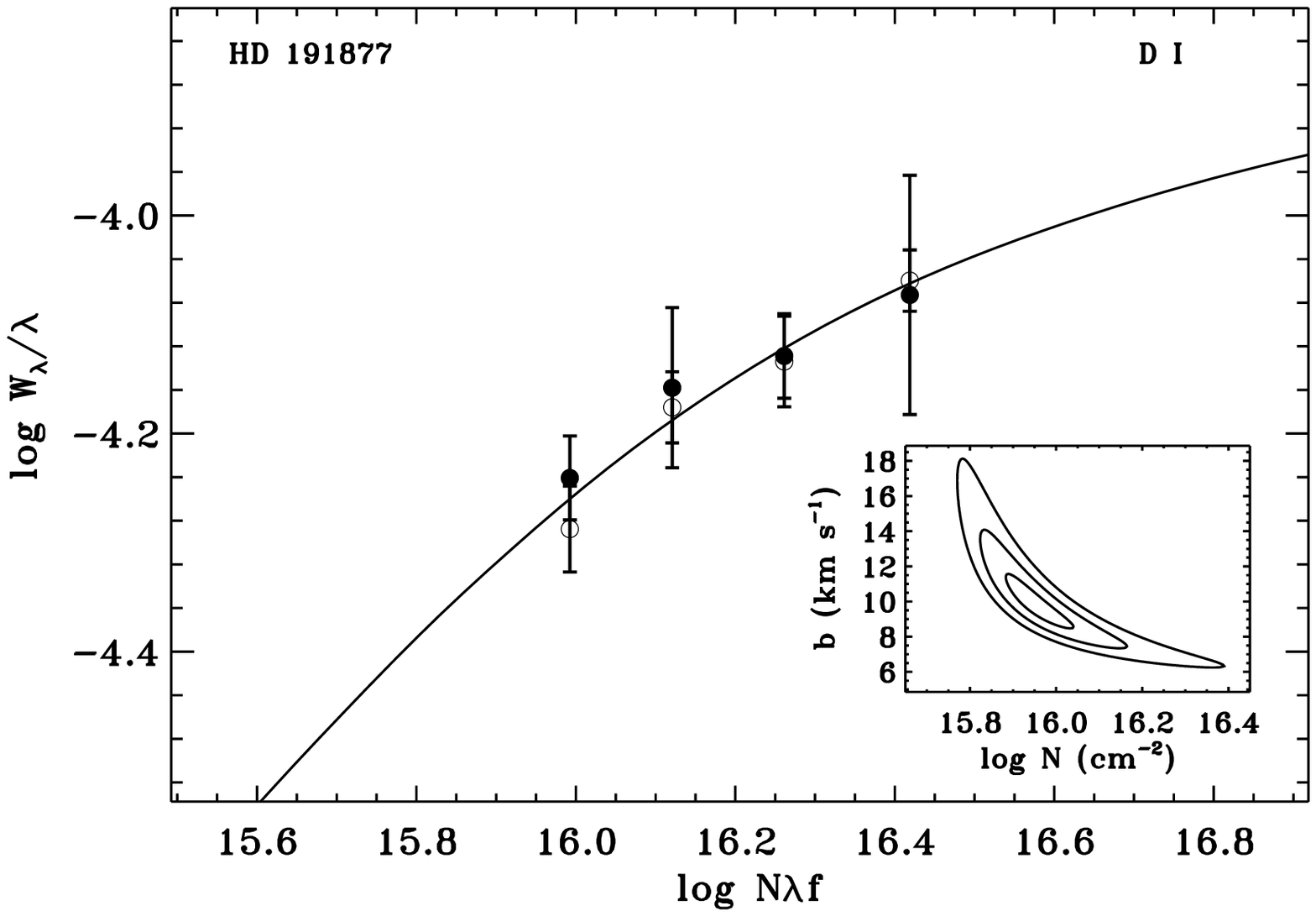}
\caption{The best fit single-component curve of growth for D I toward HD~191877, which gives log $N$(D I)=$15.95\pm^{0.21}_{0.12}$ (2$\sigma$) and $b=9.9\pm4.0$ km s$^{-1}$ (2$\sigma$). The filled circles are points measured from the SiC1 channel, and the open circles are points measured from the SiC2 channel. The inset shows a contour plot of the $\chi^2$ distribution, with the 1$\sigma$, 2$\sigma$, and 3$\sigma$ contours shown.}
\end{figure} 

\begin{figure}
\figurenum{10}
\epsscale{0.95}
\plotone{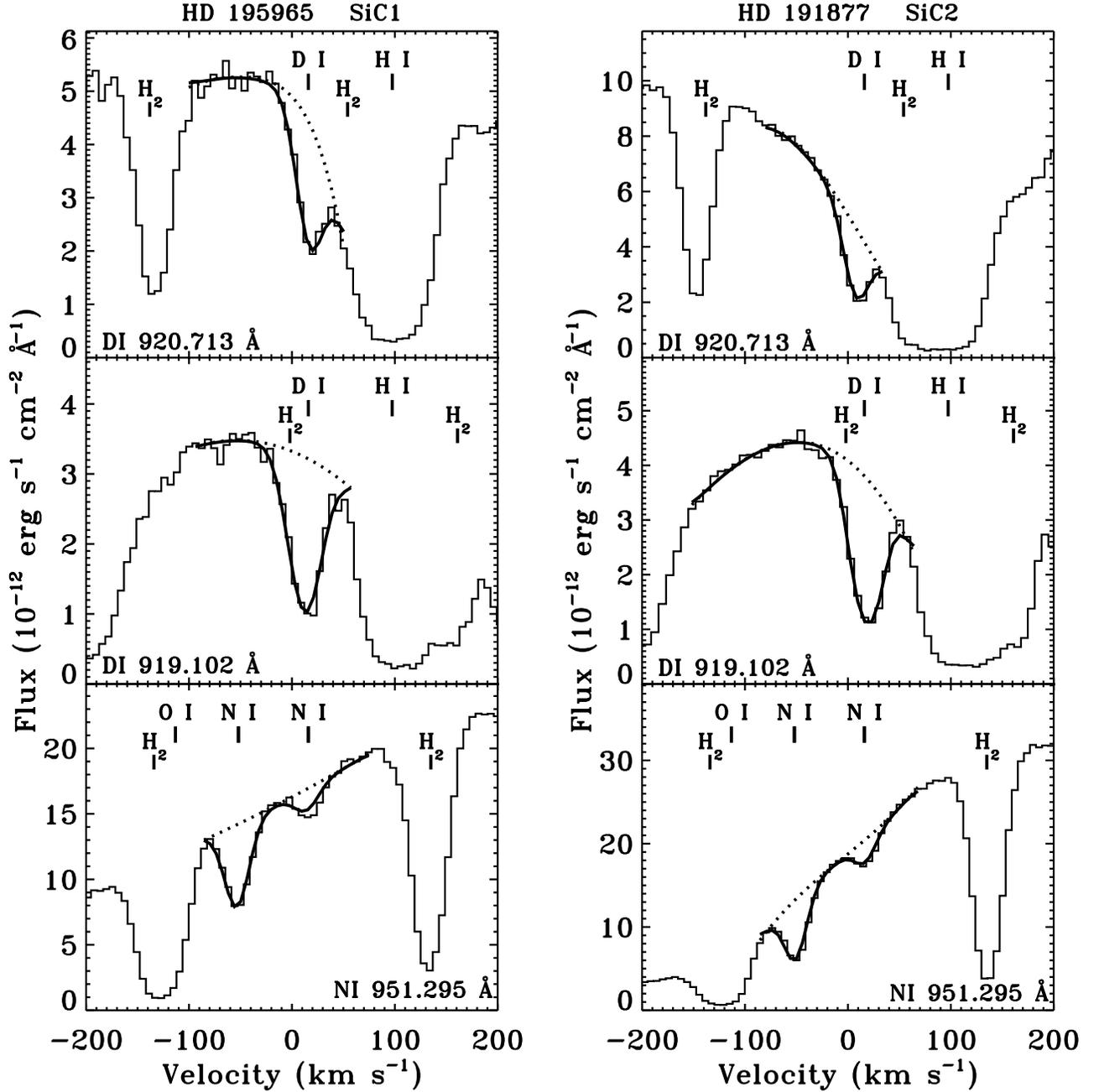}
\caption{Examples of Owens.f absorption profile fits. The histogram lines are the data, the solid thick lines are 
 the fits, and the dotted lines are the continua. Absorption features are labeled.}
\end{figure}

\begin{figure}
\figurenum{11}
\epsscale{0.55}
\plotone{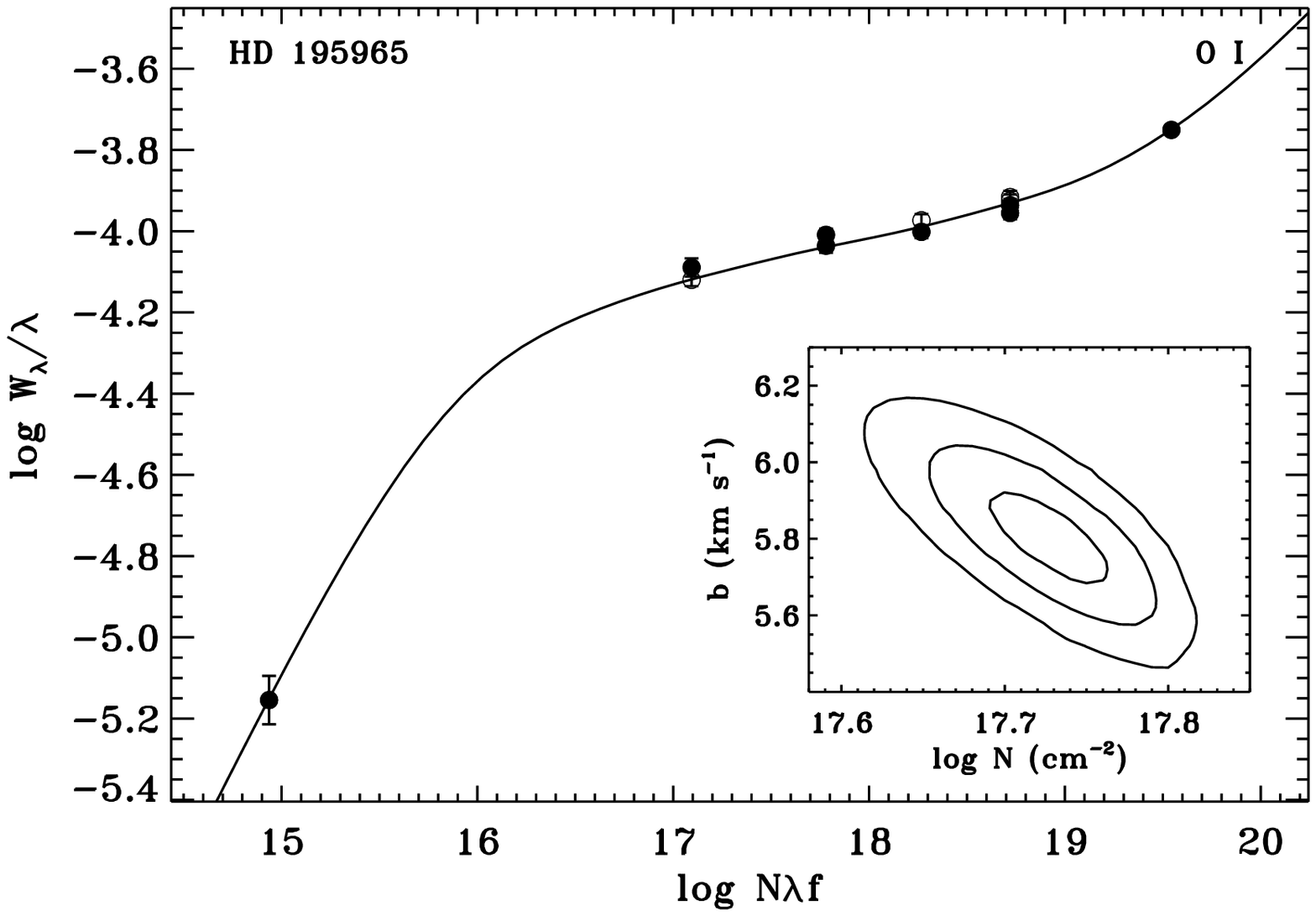}
\plotone{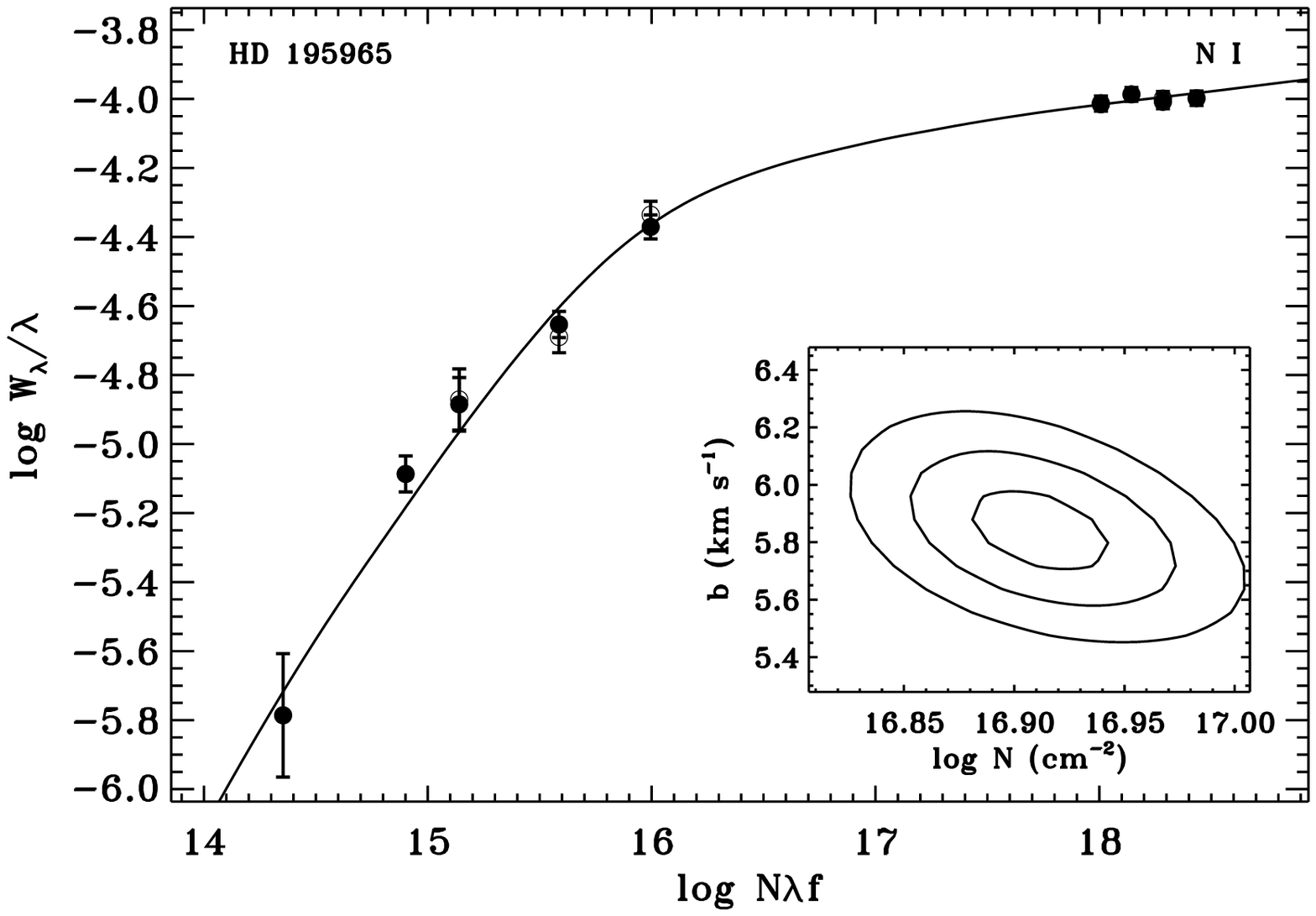}
\caption{The best fit single-component curve of growth for O I ({\it top panel}) and N I ({\it bottom panel}) toward HD~195965. The COG parameters for O I are log $N$(O I)=$17.74\pm^{0.05}_{0.09}$ (2$\sigma$) and $b=5.9\pm0.8$ km s$^{-1}$ (2$\sigma$), and the parameters for N I are log $N$(N I)=$16.92\pm^{0.05}_{0.06}$ (2$\sigma$) and $b=5.8\pm0.3$ km s$^{-1}$ (2$\sigma$). The filled circles are points measured from either the \fuse\ SiC1 or LiF1 channel or from STIS, and the open circles are points measured from either the \fuse\ SiC2 or LiF2 channel. The insets show contour plots of the $\Delta\chi^2$ distributions, with the 1$\sigma$, 2$\sigma$, and 3$\sigma$ contours shown. }
\end{figure} 

\begin{figure}
\figurenum{12}
\epsscale{0.55}
\plotone{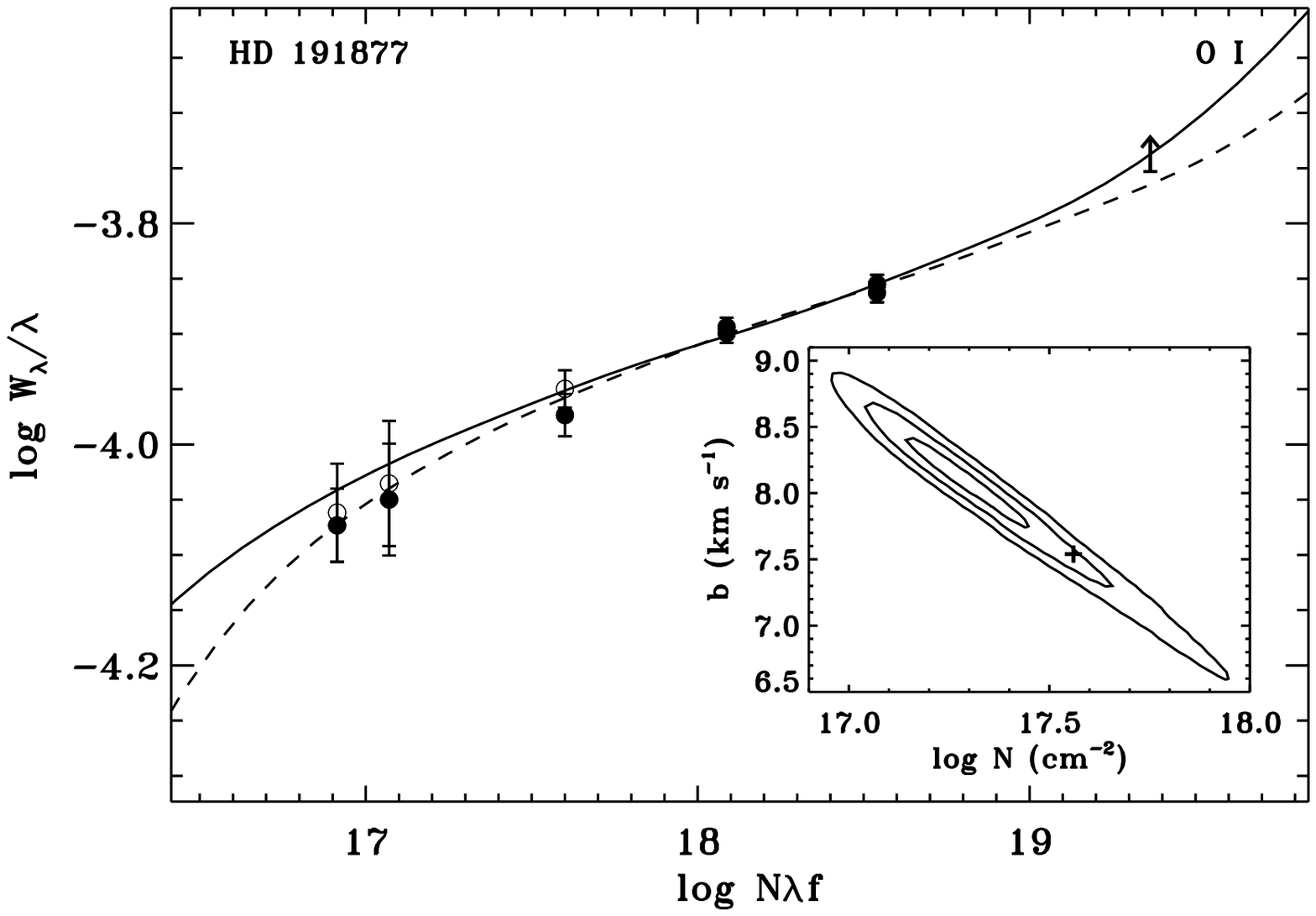}
\plotone{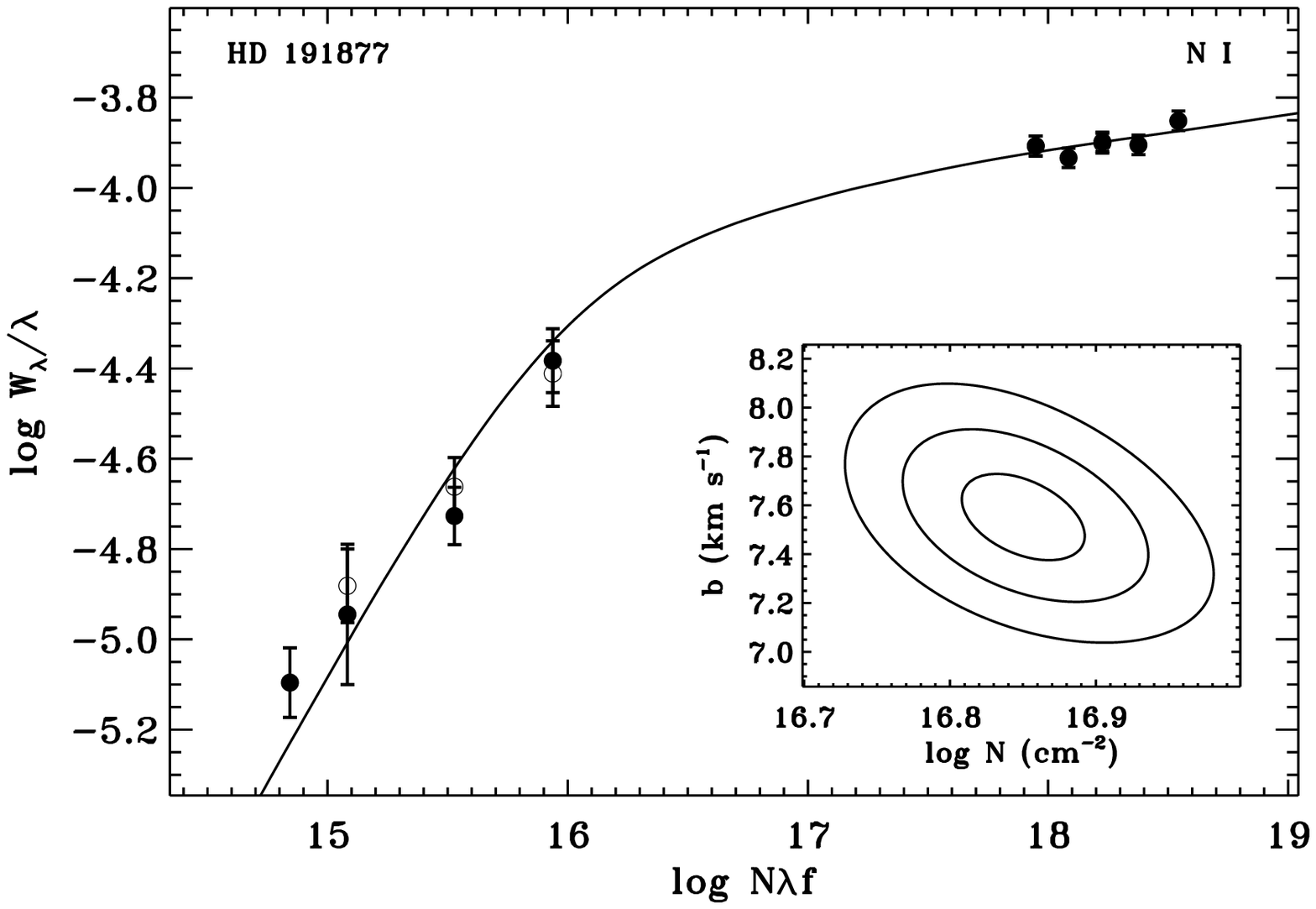}
\caption{The best fit single-component curve of growth for O I ({\it top panel}) and N I ({\it bottom panel}) toward HD~191877. The COG parameters for O I are log $N$(O I)=$17.54\pm^{0.20}_{0.12}$~cm$^{-2}$ (2$\sigma$) if $b$ is forced to be 7.5~km s$^{-1}$ as suggested by the N~I COG (the COG for these parameters is shown as the solid line in the top panel). If $b$ is allowed to vary, we find log $N$(O I)=$17.24\pm^{0.42}_{0.22}$~cm$^{-2}$ (2$\sigma$) and $b=8.2\pm^{0.05}_{0.09}$ km s$^{-1}$ (2$\sigma$), the curve for which is shown as a dashed line in the top panel. The lower limit on the 1302.169~\AA\ line (the arrow in the top panel) restricts the O~I column density to log $N$(O I)$\ge17.42$~cm$^{-2}$. The parameters for N I are log $N$(N I)=$16.85\pm^{0.08}_{0.08}$ (2$\sigma$) and $b=7.5\pm0.5$ km s$^{-1}$ (2$\sigma$). The filled circles are points measured from either the \fuse\ SiC1 or LiF1 channel, and the open circles are points measured from either the \fuse\ SiC2 or LiF2 channel. The insets show contour plots of the $\Delta\chi^2$ distributions, with the 1$\sigma$, 2$\sigma$, and 3$\sigma$ contours shown. The plus sign in the inset for O~I shows where $b$ and $N$ for the solid ($b=7.5$~km~s${-1}$) COG fall on the $\Delta\chi^2$ contours. The large error bars for some O I equivalent widths are a result of continuum placement uncertainty.}
\end{figure} 

\begin{figure}
\figurenum{13}
\epsscale{0.95}
\plotone{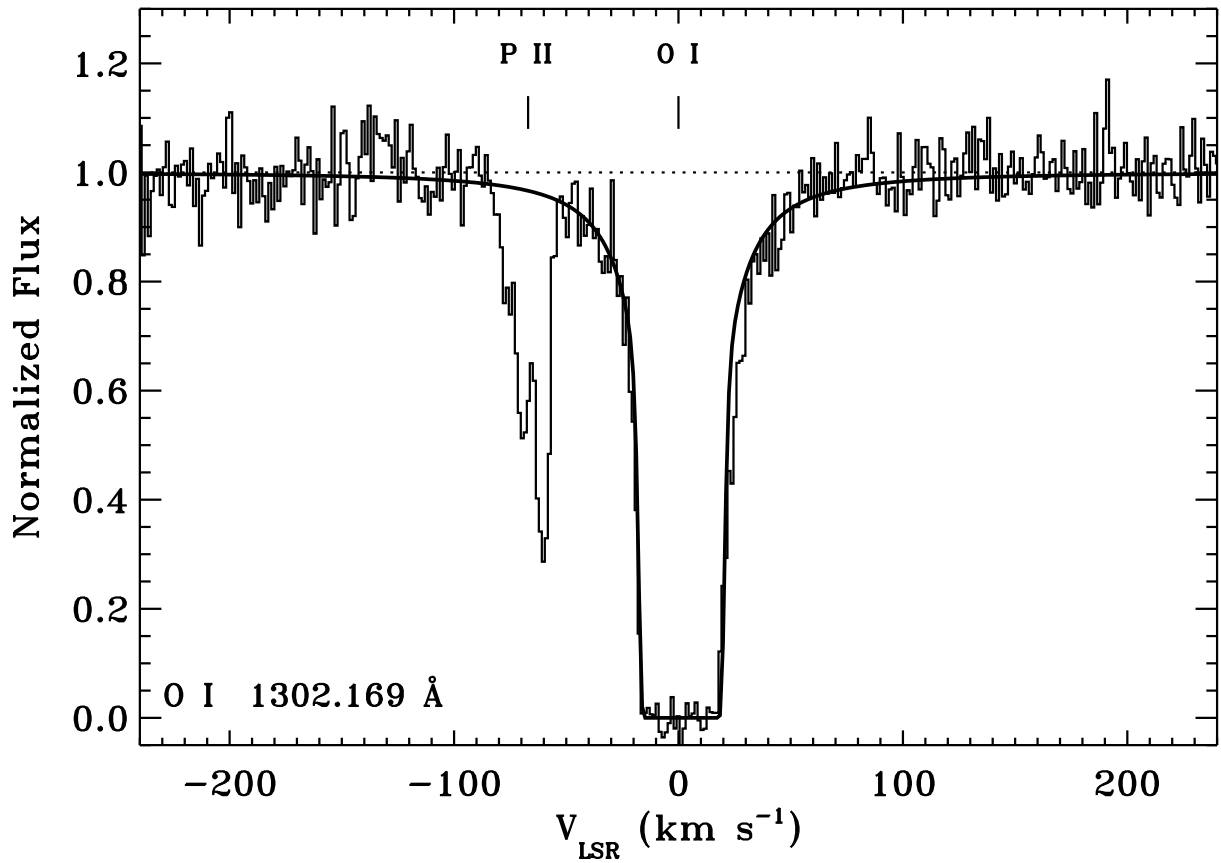}
\caption{The continuum-normalized O I 1302.169~\AA\ line profile from the STIS spectrum of HD~195965. A single-component model fit to the damping wings is also shown. The best fit model gives log $N$(O I)=17.80 and
b=6.3 \kms.}
\end{figure}

\begin{figure}
\figurenum{14}
\epsscale{0.95}
\plotone{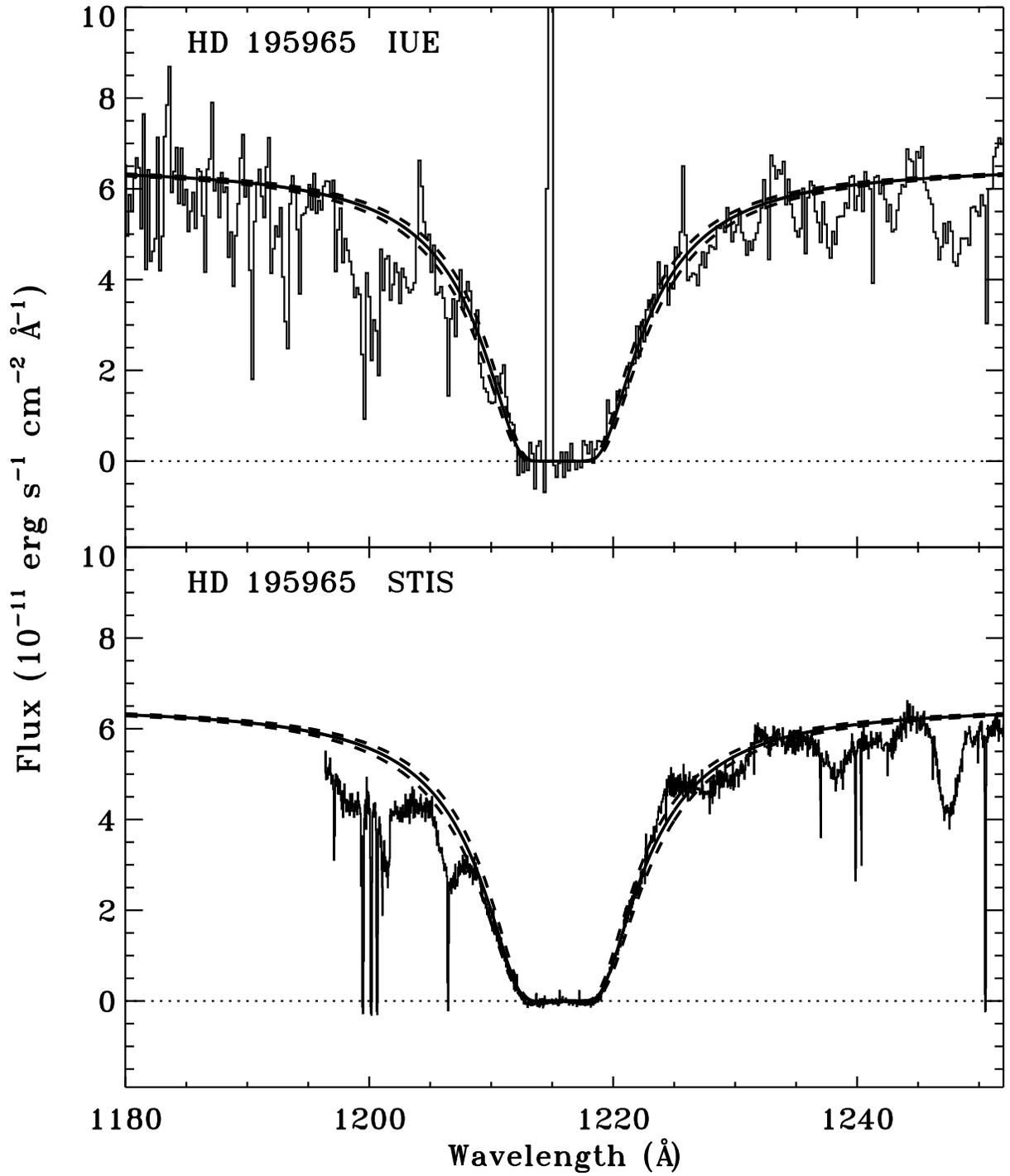}
\caption{H I Ly$\alpha$ absorption toward HD~195965. The top panel shows small-aperture \iue\ date from Diplas \& Savage (1994). The bottom panel show the STIS spectrum. The best fit Voigt profile (solid line) and the $\pm2\sigma$ profiles (dashed lines) are shown. The model shown is for log~$N$(H~I)$=20.95\pm0.05$.}
\end{figure}

\begin{figure}
\figurenum{15}
\epsscale{1.0}
\plotone{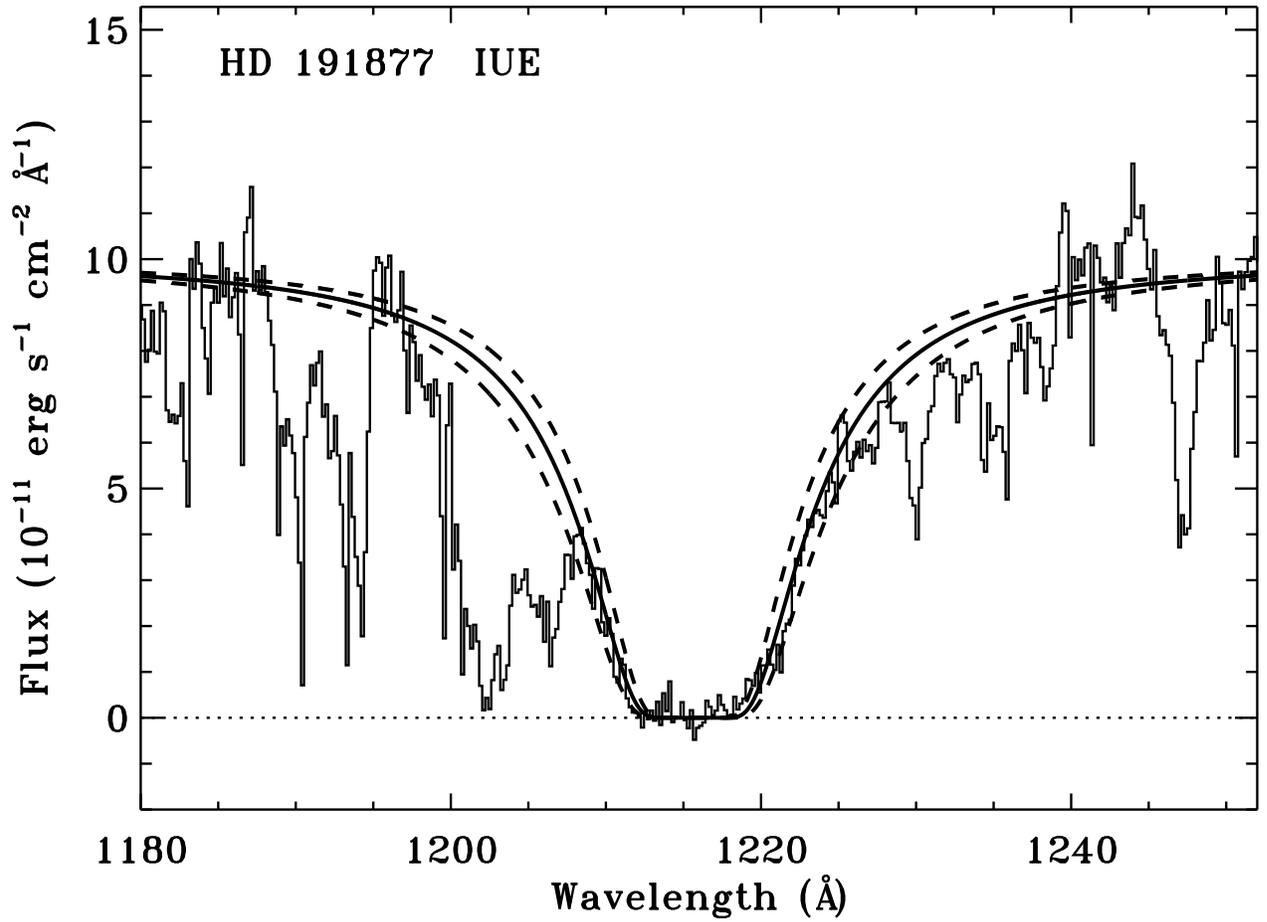}
\caption{H I Ly$\alpha$ absorption toward HD~191877. The spectrum is small-aperture \iue\ date from Diplas \& Savage (1994). The best fit Voigt profile (solid line) and the $\pm2\sigma$ profiles (dashed lines) are shown. The model shown is for log~$N$(H~I)$=21.05\pm0.10$.}
\end{figure}

\begin{figure}
\figurenum{16}
\epsscale{0.90}
\plotone{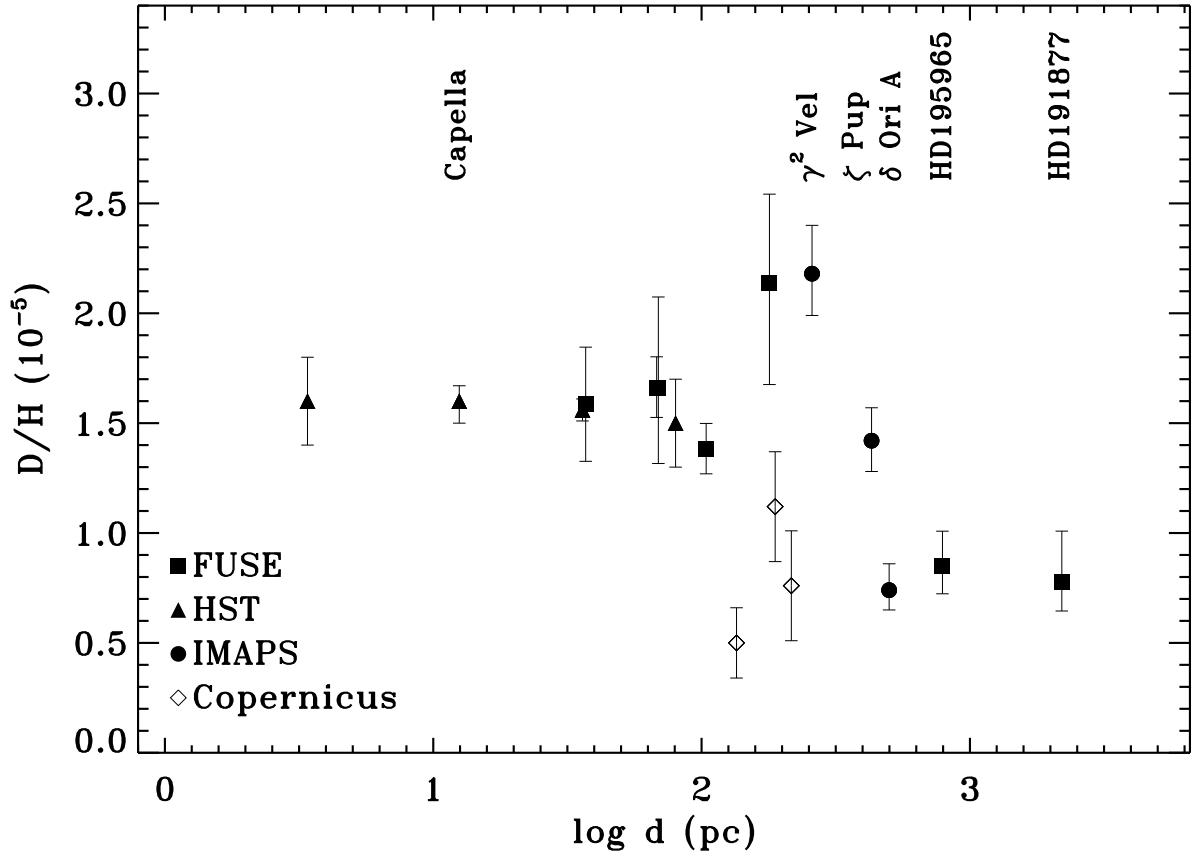}
\caption{D/H measurements for 17 sight lines, plotted versus distance to the continuum source. The sight lines studied in this paper are labeled, as are the sight lines observed by \imaps\ and the Capella sight line. Following Moos et al. (2002), the error bars shown are  2$\sigma$/2. Adapted from Moos et al. (2002).}
\end{figure}

\begin{figure}
\figurenum{17}
\epsscale{0.98}
\plotone{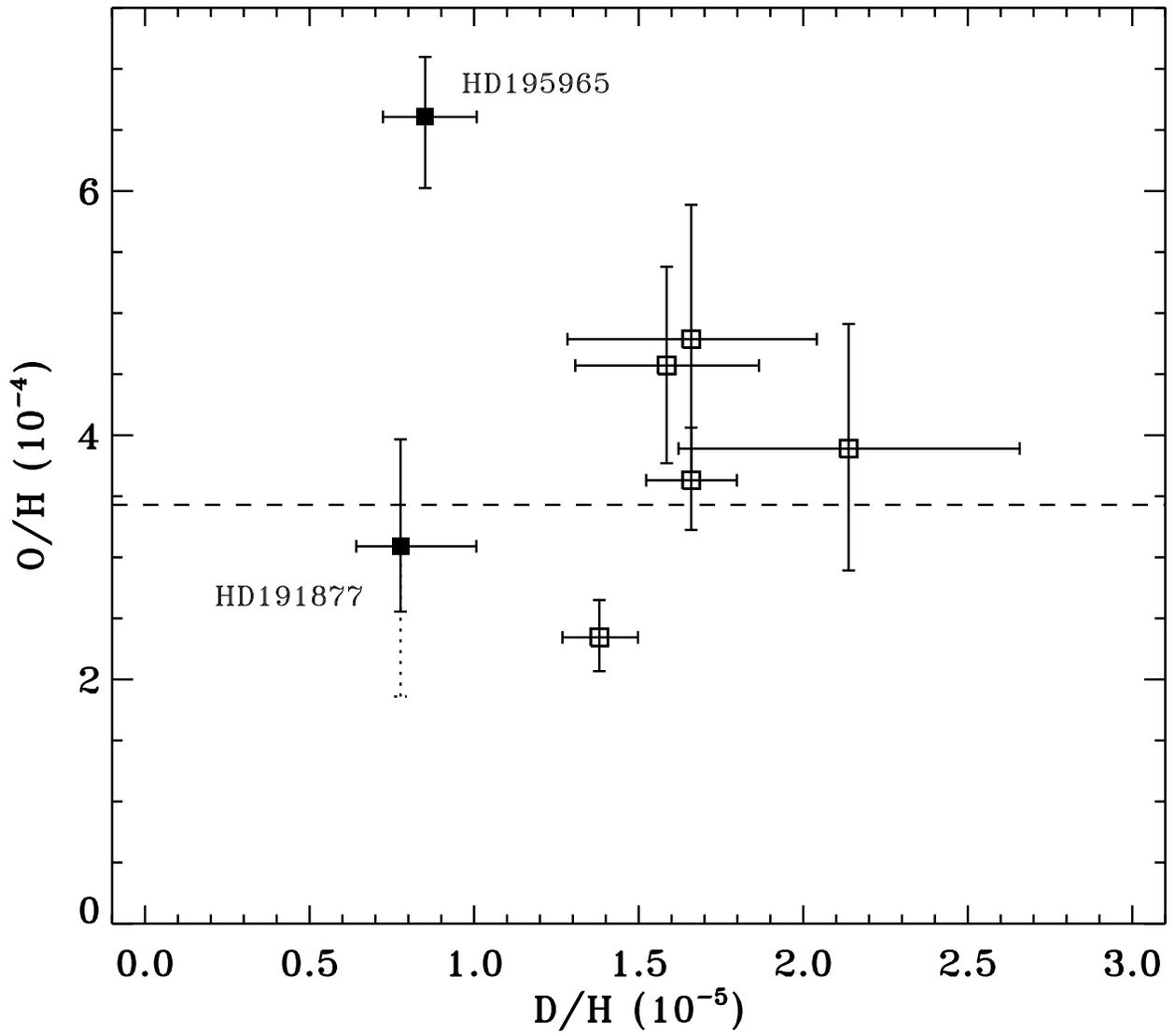}
\caption{O I/H I for HD~195965, along with five of the seven original \fuse\ sight lines with reliable H I measurements (Moos et al. 2002). Following Moos et al. (2002), the error bars shown are  2$\sigma$/2. The dashed line marks the O/H value found by Meyer et al. (1998) for the local ISM, corrected for the updated oscillator strength of the 1355.598~\AA\ line (Meyer 2001). The dotted lower error bar on the HD~191877 point show the lower limit on O/H provided by the 1302.169~\AA\ \iue\ measurement.}
\end{figure} 

\begin{figure}
\figurenum{18}
\epsscale{0.98}
\plotone{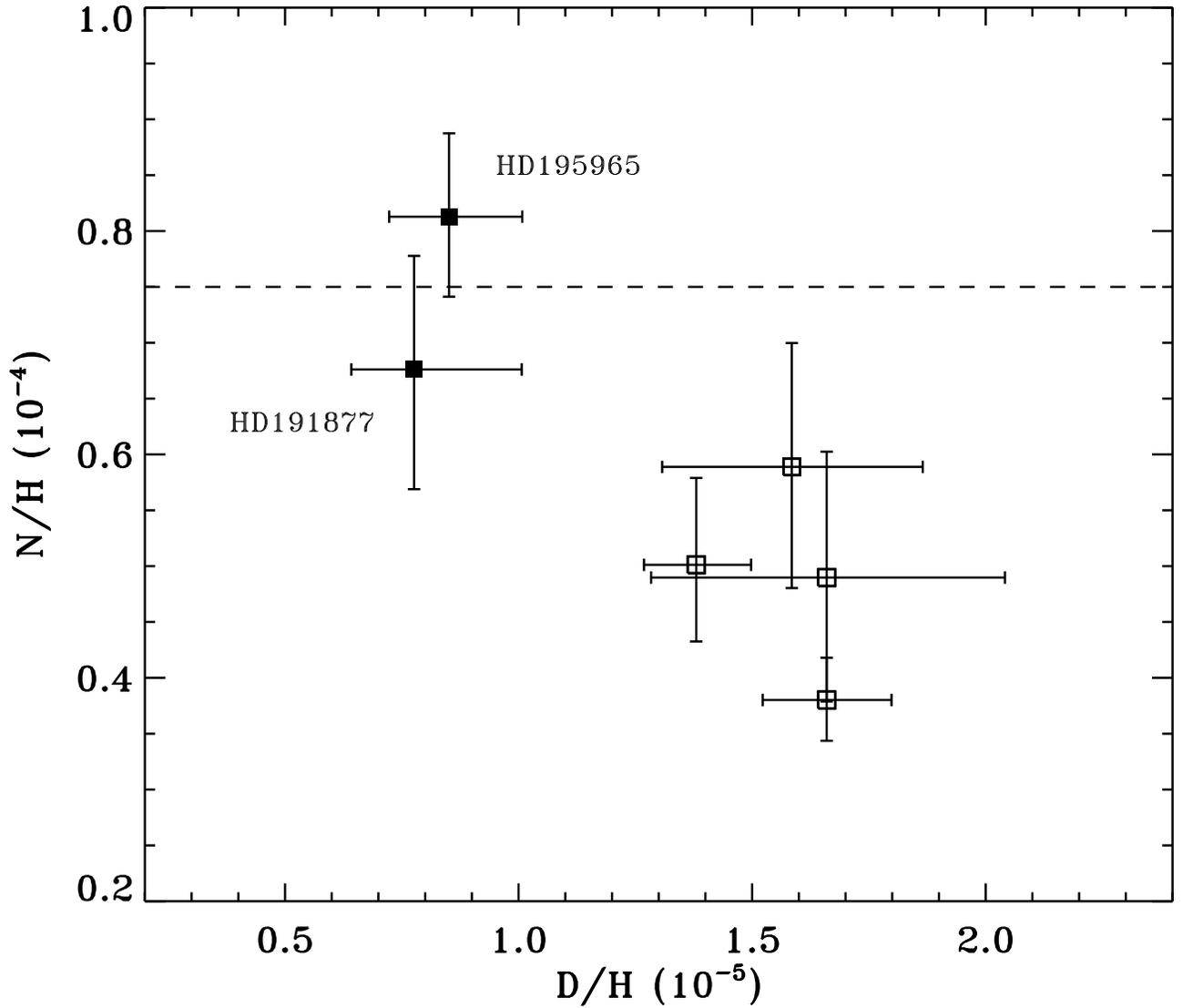}
\caption{N I/H I for HD~195965 and HD~191877, along with four of the original \fuse\ sight lines (Moos et al. 2002) for which N/H could be determined. Following Moos et al. (2002), the error bars shown are  2$\sigma$/2. The dashed line marks the N/H value found by Meyer et al. (1997) for the local ISM.}
\end{figure} 

\clearpage

\begin{deluxetable}{lcc}
\tablewidth{0pc}
\tablenum{1}
\tablecaption{Sight Line Information\tablenotemark{a}} 
\tablehead{
\colhead{Name}& \colhead{HD~195965} & \colhead{HD~191877} }
\startdata
{\it l} & 85$^{\circ}\!\!$.71 & 61$^{\circ}\!\!$.57 \\ 
{\it b} & 5$^{\circ}\!\!$.00 & -6$^{\circ}\!\!$.45 \\ 
Distance\tablenotemark{b} (pc) & $794\pm200$ & $2200\pm550$ \\ 
z dist. (pc) & 69 & -249 \\ 
V (mag) & 6.98 & 6.26 \\ 
E(B-V) & 0.25 & 0.18\\ 
Spectral Type  & B0V & B1Ib \\ 
Log [N(HI) (cm$^{-2}$)] (\iue) & 20.90 $\pm$ 0.09 & 20.90 $\pm$ 0.10 \\
\enddata
\tablenotetext{a}{From Diplas \& Savage (1994) and references therein.}
\tablenotetext{b}{We have assumed 25\% uncertainty associated with the spectroscopic parallax measurements.}
\end{deluxetable} 

\clearpage

\begin{deluxetable}{lcccccc}
\tabletypesize{\scriptsize}
\tablewidth{0pc}
\tablenum{2}
\tablecaption{Observation Log} 
\tablehead{
\colhead{Star}&\colhead{Instrument/Grating}& \colhead{Dataset ID} & \colhead{Observation Date} & \colhead{Exposure Time}& \colhead{Aperture} & \colhead{$\lambda$ coverage} \\
\colhead{}& \colhead{}& \colhead{} & \colhead{} & \colhead{(s)}& \colhead{}& \colhead{(\AA)}}
\startdata
HD~195965 & FUSE       & P1028802, P1028803        & 2000 Nov 8,9& 30000  & LWRS                      & $905-1187$\\
HD~195965 & FUSE       & P1028801\tablenotemark{a} & 2000 Jun 20 & 6400   & LWRS                      & $905-1187$\\
HD~195965 & STIS/E140H & O6BG01010                 & 2001 Oct 9  & 415    &0.1\arcs $\times$ 0.03\arcs& $1197-1401$\\
HD~195965 & IUE        & SWP10843                  & 1980 Dec 20 & 1560   & Small                     & $1150-1970$\\
HD~191877 & FUSE       & P2051101                  & 2001 Jul 30 & 28100  & LWRS                      & $905-1187$\\
HD~191877 & FUSE       & P1028701\tablenotemark{a} & 2000 Jun 5  & 6100   & LWRS                      & $905-1187$\\
HD~191877 & IUE        & SWP02837                  & 1978 Oct 2  & 2280   & Small                     & $1150-1970$\\
\enddata
\tablenotetext{a}{Datasets P1028701 and P1028801 were short snapshots. These data were included for the profile fitting analysis, but not for the curves of growth.} 
\end{deluxetable}

\begin{deluxetable}{ll}
\tabletypesize{\small}
\tablewidth{0pc}
\tablenum{3}
\tablecaption{Interfering Lines} 
\tablehead{
\colhead{$\lambda$}&\colhead{Contaminating Lines} \\
\colhead{(\AA)}& \colhead{} }
\startdata
D I 972.272 & H I Ly$\gamma$, O I $\lambda972.143$\\
D I 949.484 & H I Ly$\delta$, H$_2$ Lyman (14-0) P(2) $\lambda949.353$, H$_2$ Werner (3-0) P(2) $\lambda949.611$\\
D I 937.548 & H$_2$ Werner (4-0) P(4) $\lambda937.557$, Fe II $\lambda937.652$ \\
D I 930.495 & O I $\lambda930.257$, H$_2$ Werner (4-0) R(2) $\lambda930.446$, H$_2$ Werner (4-0) Q(1) $\lambda930.578$\\
\enddata
\end{deluxetable}

\begin{deluxetable}{lcccccc}
\tabletypesize{\small}
\tablewidth{0pc}
\tablenum{4}
\tablecaption{Interfering H$_2$ Line Removal Parameters} 
\tablehead{
\colhead{} &
\multicolumn{3}{c}{Interfering H$_2$} & \multicolumn{3}{c}{Template H$_2$} \\
\colhead{D I $\lambda$} & \colhead{ID}& \colhead{$\lambda$} & \colhead{log $\lambda f$\tablenotemark{a}} & \colhead{ID} & \colhead{$\lambda$} & \colhead{log $\lambda f$\tablenotemark{a}}\\
\colhead{(\AA)}&\colhead{}& \colhead{(\AA)} & \colhead{}&\colhead{}& \colhead{(\AA)} & \colhead{} }
\startdata
919.102 & Werner $(5-0)$ $R(4)$ & 919.047 & 0.931 & Lyman $(13-0)$ $R(4)$ & 962.150 & 0.929\\
920.713 & Werner $(5-0)$ $Q(4)$ & 920.830 & 0.0678 & Lyman $(12-0)$ $R(4)$ & 970.830 & 0.194\\
922.900 & Lyman $(18-0)$ $P(3)$ & 922.893 & -0.820 & Lyman $(17-0)$ $R(4)$ & 928.440 & -0.511\\
925.974 & none                & \nodata & \nodata  & \nodata & \nodata & \nodata  \\
\enddata
\tablenotetext{a}{Oscillator strengths from Abgrall, Roueff, \& Drira 2000.} 
\end{deluxetable}

\begin{deluxetable}{lccc}
\tablewidth{0pc}
\tablenum{5}
\tablecaption{Measured Equivalent Widths Toward HD~195965} 
\tablehead{
\colhead{Wavelength }& \colhead{log $\lambda f$\tablenotemark{a}}& \colhead{W$_{\lambda}$ (SiC1 or LiF1)\tablenotemark{b}} & \colhead{W$_{\lambda}$ (SiC2 or LiF2)\tablenotemark{b}} \\
\colhead{(\AA) }& \colhead{}& \colhead{(m\AA)} & \colhead{(m\AA)} }
\startdata
DI 919.102     & 0.0425 & $48.8\pm3.9$  & $38.4\pm4.3$   \\
DI 920.713     & 0.171 & $54.9\pm5.3$  & $64.3\pm8.4$    \\
DI 922.900     & 0.312 & $66.1\pm12.2$  & $74.2\pm3.5$    \\
DI 925.974     & 0.469 & $77.4\pm24.9$  & $76.5\pm10.2$   \\
OI 921.860     & 0.0402 & $84.9\pm3.6$  & $90.3\pm3.4$   \\
OI 922.220     & -0.646 & $75.1\pm4.2$  & $70.0\pm2.7$    \\
OI 936.630     & 0.527 & $93.3\pm3.6$  & $99.6\pm3.8$    \\
OI 1039.230    & 0.980 & $120.4\pm4.6$ & $126.1\pm4.8$   \\
OI 1039.230    & 0.980 & $115.2\pm4.4$ & $123.5\pm4.8$   \\
OI 1302.169\tablenotemark{c}    & 1.804  & $231.2\pm5.9$  & \nodata       \\
OI 1355.598\tablenotemark{c}    & -2.803 & $10.7\pm2.1$   & \nodata       \\
NI 951.079     & -0.911 & $40.5\pm3.4$  & $43.9\pm4.1$   \\
NI 951.295     & -1.766 & $12.4\pm2.4$  & $12.8\pm2.9$   \\
NI 953.415     & 1.100 & $92.0\pm4.4$  & $92.6\pm4.6$    \\
NI 953.655     & 1.377 & $93.6\pm4.6$  & $95.6\pm4.6$    \\
NI 959.494     & -1.321 & $21.3\pm1.9$  & $19.6\pm2.2$   \\
NI 1134.165\tablenotemark{d}    & 1.237 & $116.9\pm5.6$  & \nodata         \\
NI 1134.165\tablenotemark{d}    & 1.237 & $113.9\pm5.6$  & \nodata         \\
NI 1159.817\tablenotemark{d}    & -2.006 & $9.5\pm1.2$  & \nodata        \\
NI 1160.937\tablenotemark{d}    & -2.553 & $1.9\pm1.0$  & \nodata        \\
\enddata
\tablenotetext{a}{Oscillator strengths from Morton 1991, except that for the OI 1355.598~\AA\ line, which is from Welty 1999.} 
\tablenotetext{b}{Lines with $\lambda<1050$~\AA\ were measured from the SiC spectra, and lines with $\lambda>1000$~\AA\ were measured from the LiF spectra.} 
\tablenotetext{c}{These measurements are from the STIS spectrum.} 
\tablenotetext{d}{Detector x-walk effects due to the bright airglow lines at these wavelengths prevented the measurement of the line strengths in the LiF2 channel.} 
\end{deluxetable}

\begin{deluxetable}{lccc}
\tablewidth{0pc}
\tablenum{6}
\tablecaption{Measured Equivalent Widths Toward HD~191877} 
\tablehead{
\colhead{Wavelength }& \colhead{log $\lambda f$\tablenotemark{a}}& \colhead{W$_{\lambda}$ (SiC1 or LiF1)\tablenotemark{b}} & \colhead{W$_{\lambda}$ (SiC2 or LiF2)\tablenotemark{b}} \\
\colhead{(\AA) }& \colhead{}& \colhead{(m\AA)} & \colhead{(m\AA)} }
\startdata
DI 919.102     & 0.0425 & $52.8\pm4.9$   & $47.4\pm4.5$   \\
DI 920.713     & 0.171 & $64.0\pm11.8$  & $61.4\pm4.8$    \\
DI 922.900     & 0.312 & $68.6\pm6.4$   & $67.8\pm6.8$    \\
DI 925.974     & 0.469 & $78.3\pm22.5$  & $80.7\pm5.4$    \\
OI 921.860     & 0.0402 & $98.0\pm4.4$   & $103.5\pm4.1$  \\
OI 922.220     & -0.646 & $77.9\pm6.2$   & $80.0\pm8.6$   \\
OI 925.446     & -0.490 & $82.5\pm10.2$  & $85.3\pm11.9$   \\
OI 936.630     & 0.527 & $119.6\pm2.4$  & $118.1\pm2.4$   \\
OI 1039.230    & 0.980 & $145.0\pm2.9$  & $145.1\pm2.9$   \\
OI 1039.230    & 0.980 & $142.6\pm2.9$  & $142.6\pm2.9$   \\
NI 951.079     & -0.911 & $39.4\pm7.0$   & $36.9\pm6.7$   \\
NI 951.295     & -1.766 & $10.8\pm4.6$   & $12.5\pm2.6$   \\
NI 953.415     & 1.100 & $118.1\pm6.2$  & $118.1\pm6.2$   \\
NI 953.655     & 1.377 & $120.6\pm6.2$  & $119.8\pm6.2$   \\
NI 959.494     & -2.006 & $18.0\pm2.8$   & $20.9\pm3.4$   \\
NI 1159.817\tablenotemark{c}    & -2.553 & $9.3\pm1.8$    & \nodata        \\
NI 1134.165\tablenotemark{c}    & 1.237 & $132.2\pm6.7$  & \nodata         \\
NI 1134.415\tablenotemark{c}    & 1.528 & $141.3\pm7.2$  & \nodata         \\
NI 1134.980\tablenotemark{c}    & 1.693 & $159.8\pm8.3$  & \nodata         \\
\enddata
\tablenotetext{a}{Oscillator strengths from Morton 1991.} 
\tablenotetext{b}{Lines with $\lambda<1050$~\AA\ were measured from the SiC spectra, and lines with $\lambda>1000$~\AA\ were measured from the LiF spectra. The O I 1039~\AA\ lines was measured in both LiF and SiC channels.} 
\tablenotetext{c}{Detector x-walk effects due to the bright airglow lines at these wavelengths prevented the measurement of the line strengths in the LiF2 channel.} 
\end{deluxetable}

\begin{deluxetable}{lccccccc}
\tabletypesize{\scriptsize}
\tablewidth{0pc}
\tablenum{7}
\tablecaption{Column Densities\tablenotemark{a}} 
\tablehead{
\colhead{}& \multicolumn{3}{c}{HD~195965} & \multicolumn{2}{c}{HD~191877} \\
\colhead{Name} & \colhead{COG} & \colhead{Profile fit} & \colhead{Adopted Value\tablenotemark{b}} & \colhead{COG} & \colhead{Profile fit} & \colhead{Adopted Value\tablenotemark{b}}}
\startdata
log N(D I) & $15.83\pm^{0.14}_{0.11}$ & $15.97\pm^{0.13}_{0.13}$ & $15.88\pm^{0.14}_{0.13}$ & $15.95\pm^{0.21}_{0.12}$  & $15.93\pm^{0.19}_{0.11}$ & $15.94\pm^{0.21}_{0.12}$  \\
log N(H I)\tablenotemark{c}     & \nodata & $20.95\pm0.05$ & $20.95\pm0.05$   & \nodata & $21.05\pm0.10$ & $21.05\pm0.10$ \\
log N(O I)\tablenotemark{d} & $17.74\pm^{0.05}_{0.09}$ & $17.80\pm^{0.05}_{0.09}$ & $17.77\pm^{0.04}_{0.06}$ & $17.54\pm^{0.20}_{0.12}$\tablenotemark{e} & \nodata & $17.54\pm^{0.20}_{0.12}$\tablenotemark{e}\\
log N(N I) & $16.92\pm^{0.05}_{0.06}$ & $16.83\pm^{0.04}_{0.04}$ & $16.85\pm^{0.07}_{0.06}$ & $16.85\pm^{0.08}_{0.08}$ & $16.93\pm^{0.07}_{0.10}$ & $16.88\pm^{0.08}_{0.10}$ \\
\enddata
\tablenotetext{a}{Error bars are $2\sigma$ estimates (95\% confidence). For all of the adopted values except O I the uncertainties were chosen to be the largest uncertainties of the individual values, because the two measurements were made on the same data and are therefore not independent. For O I toward HD~195965 the measurements are largely independent, so the errors were added in quadrature.}
\tablenotetext{b}{For D I and N I, the adopted value is the weighted mean of the single-component curve of growth and profile fitting results. When only one method was available ({\it e.g.,} H I), the result from that method is the adopted value.}
\tablenotetext{c}{The H I column densities were derived by model fits to the damping wings of Ly$\alpha$. For HD 195965, Ly$\alpha$ was observed with STIS, and for HD~191877 we reanalyzed \iue\ data from Diplas \& Savage 1994.}
\tablenotetext{d}{The O I column density toward HD~195965 was derived by a single-component curve of growth and a model fit to the damping wings of the 1302.169 \AA\ line in the STIS data.}
\tablenotetext{e}{The O I column density listed in the table is the value derived when $b$ is set to 7.5~km~s$^{-1}$. If $b$ is allowed to vary, the best fit COG gives log~$N$(O~I)=$17.24\pm^{0.42}_{0.22}$~cm$^{-2}$ (see discussion in \S~5).}
\end{deluxetable} 

\begin{deluxetable}{lccc}
\tablewidth{0pc}
\tablenum{8}
\tablecaption{Column Density Ratios\tablenotemark{a}} 
\tablehead{
\colhead{Name}& \colhead{HD~195965} & \colhead{HD~191877} & \colhead{Local ISM\tablenotemark{b} }}
\startdata
D/H	       & $(0.85\pm^{0.34}_{0.24})\times10^{-5}$ & $(0.78\pm^{0.52}_{0.25})\times10^{-5}$ & $(1.52\pm0.15)\times10^{-5}$ \\
D/O	       & $(1.29\pm^{0.51}_{0.37})\times10^{-2}$ & $(2.51\pm^{2.14}_{0.86})\times10^{-2}$\tablenotemark{c} & $(3.99\pm0.38)\times10^{-2}$ \\
D/N	       & $(1.07\pm^{0.45}_{0.31})\times10^{-1}$ & $(1.15\pm^{0.75}_{0.36})\times10^{-1}$ & $(3.30\pm0.40)\times10^{-1}$ \\
O/H	       & $(6.61\pm^{1.03}_{1.11})\times10^{-4}$ & $(3.09\pm^{1.98}_{0.98})\times10^{-4}$\tablenotemark{c} & $(3.03\pm0.42)\times10^{-4}$ \\
N/H	       & $(7.94\pm^{1.69}_{1.34})\times10^{-5}$ & $(6.76\pm^{2.22}_{1.97})\times10^{-5}$ & $(4.24\pm0.62)\times10^{-5}$ \\
O/N	       & $8.32\pm^{1.66}_{1.52}$ & $4.57\pm^{2.83}_{1.45}$ & $8.1\pm0.8$ \\
\enddata
\tablenotetext{a}{Error bars are 2$\sigma$ estimates (95\% confidence).}
\tablenotetext{b}{Mean Local ISM values from Moos et al. 2002 and references therein. The uncertainties for these values are $2\sigma$ in the mean.}
\tablenotetext{c}{The D/O and O/H ratios toward HD~191877 listed in the table use the O~I column density found by setting $b$ to 7.5~km~s$^{-1}$. If $b$ is allowed to vary, we find D/O=$(5.01\pm^{8.74}_{2.33})\times10^{-2}$, O/H=$(1.55\pm^{2.56}_{0.69})\times10^{-4}$ (see discussion in \S~5).}
\end{deluxetable} 

\end{document}